\documentclass[%
reprint,
amsmath,amssymb,
prb,
]{revtex4-2}

\usepackage{graphicx}
\usepackage{dcolumn}
\usepackage{bm}
\usepackage{braket}
\usepackage{dsfont}
\usepackage{xcolor}
\usepackage{tikz-cd}
\usepackage{multirow}

\DeclareMathAlphabet\mathbfcal{OMS}{cmsy}{b}{n}

\numberwithin{equation}{section}

\newcommand{\abs}[1]{\left\vert#1\right\vert}

\renewcommand{\i}{\mathrm{i}}
\newcommand{\tr}{\textrm{Tr}\;}

\newcommand{\1}{\mathds{1}}

\newcommand{\ubar}[1]{\underline{#1}}

\renewcommand{\H}{\mathcal{H}}

\newcommand{\N}{\mathcal{N}}

\newcommand{\Lat}{\bar{\mathbb{Z}}^D}
\newcommand{\Lattwo}{\bar{\mathbb{Z}}^2}
\newcommand{\Sp}{\mathbf{S}}

\def \q{\mathbf{q}}

\begin{document}

\title{Local correlations in partially  dual-unitary lattice models}

\author{Niclas Krieger}
\affiliation{Duisburg-Essen University, Duisburg, Germany}
\author{Vladimir Al. Osipov}
\affiliation{The Institute for Advanced Study in Mathematics, Harbin Institute of Technology, 92 West Da Zhi Street, Harbin 150001, China}
\email{Vladimir.Al.Osipov@gmail.com}
\author{Boris Gutkin}
\affiliation{H.I.T.-Holon Institute of Technology, 52 Golomb Street, POB 305 Holon 5810201, Israel}
\author{Thomas Guhr}
\affiliation{Duisburg-Essen University, Duisburg, Germany}

\date{\today}

\begin{abstract}
We consider the problem of local correlations in the kicked, dual-unitary   coupled  maps on $D$-dimensional lattices. We demonstrate that for $D\geq2$,   fully dual-unitary systems exhibit  ultra-local correlations: the correlations between any pair of operators with a local support vanish in a finite number of time steps. In addition, for $D=2$,  we consider the partially dual-unitary regime of the model, where  the dual-unitarity  applies to only one of the two spatial directions. For this case, we show that correlations generically decay exponentially  and provide an explicit formula for the correlation function between the operators supported on two and  four neighbouring sites.          
\end{abstract}



\maketitle

\section{Introduction }

Until recently, the vast majority of research in the field of quantum chaos has been limited to systems with few degrees of freedom, even though chaotic spectral statistics were first found in the atomic nuclear, which essentially is a many-body problem~\cite{haake, guhr}.  Indeed, the many-body quantum systems, where the Hilbert space dimension grows exponentially with the number of degrees of freedom, represent a significant challenge for numerical and analytical studies. In recent years substantial progress in the field has been achieved due to the introduction of new classes of many-body models and the development of appropriate mathematical methods for their investigation. This is closely connected to a burst of activities in the field of quantum circuits, see a recent review in \cite{fisher2023random}. In this article, our attention is focused on the calculation of the correlations between localized quantum observables in dual unitary quantum systems of arbitrary dimensions, with a particular focus on $D=2$ dimensions. Dual unitary models possess a remarkable property -- their dynamics are invariant under the exchange of spatial and temporal degrees of freedom. A toy model with such property, a  chain of linearly coupled Arnold's cat maps,  was first introduced in~\cite{GutOsi15}   and subsequently studied in a number of works~\cite{GHJSC16, LiangCvit20022, Fouxon_Gutkin_2022} both on classical and quantum levels. Other examples of dual unitary models were found among different classes of systems, e.g., kicked Ising spin chain and its generalizations~\cite{AWGG16, BeKoPr18, BeKoPr19-1, GBAWG20} and circuit lattices~\cite{BeKoPr19-4}. Although their full characterisation is still absent, dual unitary models are generic and can be constructed in a systematic way~\cite{Arul19}.

In the field of many-body quantum chaos the dual unitary models attract considerable attention \cite{BeKoPr19-4, AWGG16, GopLam19, BeKoPr19-1, LakshPal2018, BeKoPr18, BWAGG19, BeKoPrPi19, BeKoPr2019operator, KPBBPT_2021,zhou2019entanglement, AVAN2016, Karl15, Arul19, Arul2021} due to their intriguing properties. On one hand, they demonstrate quantum properties akin to those of maximally chaotic many-body systems, such as Wigner-Dyson spectral statistics and insusceptibility to many-body localization effects \cite{BeKoPr18, BWAGG19, PhysRevResearch.3.023118}. On the other hand, dual-unitary models are amenable to exact treatment. In particular,   due to the combination of duality and causality, the local two-point correlation functions in these systems can be calculated exactly~\cite{BeKoPr19-4, CL20, GBAWG20}. Correlations continue to find considerable interest~\cite{borsi2022construction, ippoliti2021postselection, claeys2022exact, KPBBPT_2021, krajnik2020kardar, lu2021spacetime, prosen2021many}. Other recent research directions include aspects of matrix product states~\cite{piroli2020exact,lerose2021influence}, steady states as well as eigenstate thermalization~\cite{ippoliti2022fractal, fritzsch2021eigenstate}, computational aspects~\cite{suzuki2022computational} and random matrix statistics~\cite{flack2020statistics, bertini2021random}.

Previous studies have primarily focused on dual unitary models with a single spatial dimension. Extensions to two spatial dimensions are rare, but in~\cite{jonay2021triunitary} a ``tri\-unitarity'' model (corresponding to $D=2$) for a general class of quantum circuits was considered. In the work, we extend our investigation to encompass coupled map lattices of an arbitrary dimension $D$. Our findings reveal that starting from $D=2$ onwards, the correlations within a system exhibiting the complete spatiotemporal symmetry demonstrate an ultra-local behaviour, which implies that the correlations between operators with local support vanish identically after a finite time. In particular, this suggests that in the case $D\ge 2$, the requirement of the complete duality can be relaxed without compromising the solvability of the model. Accordingly, in the main body of this paper, we consider partially dual unitary map lattices, where the system remains invariant under the exchange of the time variable and only one of several spatial coordinates. The way of derivation is related to the one suggested in Ref.~\cite {GBAWG20}. Our central result is the explicit expression for the local correlation function in the partially dual unitary models. 
 
\section{The Main Ideas\label{main_ideas}}

The general model considered in this paper is defined for a finite piece of the $D$-dimensional lattice $ \mathbb{Z}^D$. Specifically,  let  $\Lat$  be a finite-size hyper rectangular subset of $\mathbb{Z}^D$  where  $N_i$ (for $i=1,\dots, D$) represents the number of sites along the $i$-th spatial direction so that the product  $\mathcal{N}=N_1\cdot N_2\cdot {\dots}\cdot N_D$ is the total number of sites of the lattice  $\Lat$. The unitary Floquet evolution operator $U$ acts in discrete time steps in the Hilbert space $\H^{\otimes \N}$, which is the tensor product of $\N$ local $L$-dimensional spaces $\H=\mathbb{C}^L$. In the following, we assume that the time evolution $U$ is dual unitary (at least for one spatial direction) with a unit speed of interaction propagation. Detailed information regarding the construction of $U$ with the necessary properties will be provided in the main body of the paper. 

The main object of our consideration is the reduced correlation function between two local observables, $\hat{Q}_1$ and $\hat{Q}_2$ after $t$ time steps of the evolution:
\begin{equation}\label{connectedCorr}
C(\bm r,t)=\braket{\hat{Q}_1(0)\hat{Q}_2(t)} -
\braket{\hat{Q}_1(0)}\braket{\hat{Q}_2(t)},
\end{equation}
where the evolution is provided by the action of a unitary operator $U$,
\begin{equation}\label{Corr}
 \hat{Q}_i(t)=U^{-t} Q_i  U^t, \qquad i = {1,2}. 
\end{equation}
It's important to note that the value of $t$  should be smaller than any spatial dimension of $\Lat$, ensuring that the resulting correlation function remains independent of both the size of $\Lat$ and the boundary conditions. To make our explanations more transparent we assume here that two observables, $Q_1$ and $Q_2$ are strictly local. In other words, they are localized at single lattice points $\bm{ r}_1\in \Lat$, and  $\bm{ r}_2\in \Lat$, respectively. Furthermore, since our model is shift-invariant, the correlation function depends solely on the difference $\bm r=\bm{ r}_2-\bm{ r}_1$. Consequently, we can set, without a loss of generality, that $Q_1$ is localized at the origin $\bm 0$ and $Q_2$ at the position $\bm r$, respectively. The average in eq.~\eqref{connectedCorr} is defined by the operator trace taken over the entire many-body Hilbert space: $\langle\cdot\rangle = L^{-D} \tr (\cdot)$. 

To explain the main ideas of the paper we will now consider the cases of one-dimensional ($D=1$) and two-dimensional lattices ($D=2$).  

\subsection{One-dimensional lattice (chain) of quantum maps}
\begin{figure} 
\hspace{-0.45\linewidth}
a)\includegraphics[width=0.9\linewidth]{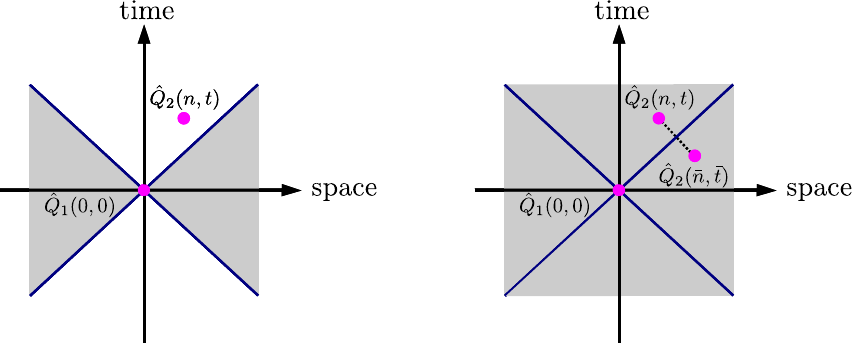}
\hspace{-0.5\linewidth} b)
\caption{\small  Representation of the connected part of the correlation function $C(n,t)$ (eq.~\ref{connectedCorr}) between two observables located at the coordinates $(0,0)$ and $(n,t)$  of the space-time grid:  (a) In the general case due to causality the correlations vanish inside the light cones defined by the inequality $|n|>|t|$ (grey regions); (b) In the case of the dual unitary model, when eq.~(\ref{DualityQ1Q2}) holds, solely the correlations along the light cone edges (dark blue lines $|n|=|t|$) do not vanish.}\label{Bild13}
\end{figure}

For a one-dimensional lattice (chain) $\bar{\mathbb{Z}}^1$, the location of an observable is determined by an integer number $n$, $\bm r=n$. Since the speed of interaction propagation equals one, the causality implies that the correlation function of many-body operators \eqref{connectedCorr} vanishes outside the light cone $|t|<|n|$, see fig.~\ref{Bild13}a. 
Furthermore, for dual unitary $U$, the correlation function remains invariant under exchange of time $t$ and the spatial coordinate $n$: 
\begin{equation}
    C(n,t)=C(t,n).\label{DualityQ1Q2}
\end{equation}
Therefore,  the correlation function $C(n,t)$ nullifies also inside the light cone $|n|<|t|$, see fig.~\ref{Bild13}b. As a result, the light cone edges $|t|=|n|$ remain the only possible location of the space-time manifold where the non-trivial correlations can arise. From a technical point of view, the calculation of the correlation function in the dual unitary case reduces to the calculation of the correlations propagating along the light cone edges, which in turn can be expressed through an expectation value of a product of transfer matrices of a reduced dimension, independent on $D$, see \cite{BeKoPr19-4,  CL20, GBAWG20}.
 
\subsection{Two-dimensional lattice of quantum maps}
\begin{figure}	
	\hspace{-0.45\linewidth}
a)\includegraphics[width=.9\linewidth]{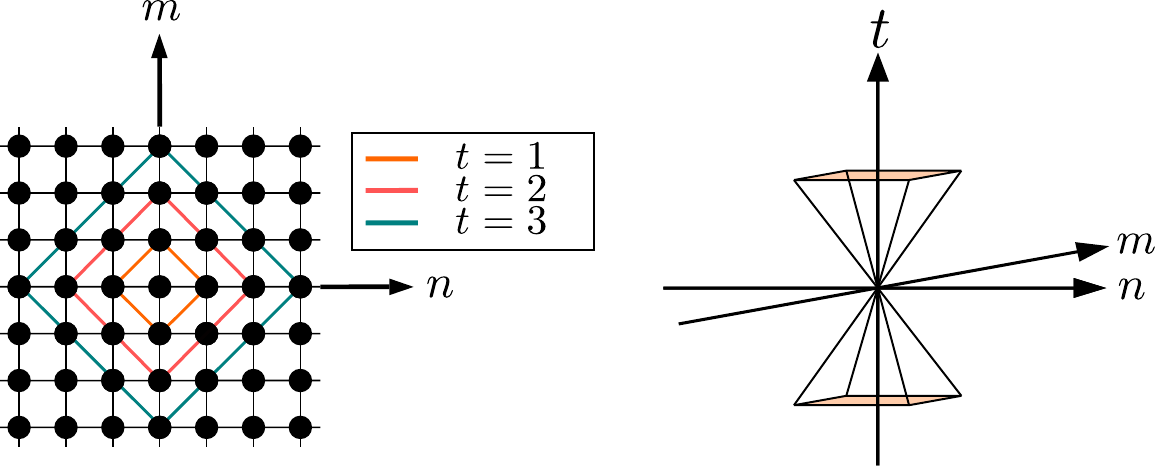}
\hspace{-0.5\linewidth} b)
	\caption{\small  On the left (a) is shown a lattice $\Lat$ with a central point located at the origin $(n=0,m=0)$ of the system. The lines $t=1$ (orange), $t=2$ (red) and $t=3$ (blue) mark the event horizon of the light cone. The observables, which are outside the cone at a given time step $t$, do not correlate with the observable at the origin. The right figure shows the entire light cone with the boundaries of the event horizon represented by the eight triangular areas.}
	\label{Bild32}
\end{figure}

In the case of a two-dimension lattice $\bar{\mathbb{Z}}^2$ the situation is somewhat different. Here any point of the lattice is labelled by a pair of integers $\bm r =(m,n)$. Due to  causality, the correlations  vanish  outside the light cone domain
\begin{equation}
    |t|\geq |n|+|m|, \label{ineq10}
\end{equation}
see fig.~\ref{Bild32}.
For systems with full spatiotemporal symmetry, the correlation function remains invariant under the exchange of $t$ and $n$, as well as under the exchange of $t$ and $m$. This implies that for the non-trivial correlations  the inequalities 
\begin{equation}
    |m|\geq|n|+|t|, \qquad |n|\geq|t|+|m| \label{ineq20}
\end{equation} 
must hold, as well. It is straightforward to see that the only point satisfying all three inequalities (\ref{ineq10}, \ref{ineq20}) is the origin of the space-time grid $t=m=n=0$. In other words, all correlations of strictly local operators vanish for any time $t>0$. A similar consideration for the operators supported on a finite number $\ell$ of sites shows that $C(\bm r, t)=0$ for $t>\ell$. For this reason, we refer to such behaviour as ultra-local correlations.

Suppose now that our system belongs to a class of partially dual unitary systems. It is invariant under the exchange of only one coordinate e.g., $n$ and time $t$. In such a case the domain of non-trivial correlations is given by  
\begin{equation}
    |t|\geq|n|+|m|\quad \cap \quad  |n|\geq|t|+|m|. \label{ineq3}
\end{equation} 
The correlation function $C(\bm r, t)$ does not necessarily vanish along the line $m=0$, $|n|=|t|$. In the body of the text, we show, similarly to the one-dimensional case, that the correlations along this line can be expressed through an expectation value of a transfer operator powers. In particular, the transfer operator eigenvalues determine the decay rates of the correlations. 
 
\subsection{Outline of the article}
In the next section, we formulate the general model of coupled  quantum maps with periodic kicks  in $D$ spatial dimensions. In this context,  we introduce the notion of partial and full dual-unitarity. In section~\ref{Sec4} we briefly recall the main results for $D=1$ case. We then introduce the correlation function for the $D=2$ model and demonstrate that, similar to the one-dimensional case, it can be expressed in the form of a three-dimensional partition function for a classical spin model. In Section~\ref{Sec5}, we derive the contraction rules that enable us to compute the correlation function.  The general expressions for the transfer operator and the correlation function are derived in section~\ref{Sec6}.  In section~\ref{Sec7} we apply our results to the model of  coupled  cat maps and kicked Ising spin lattice.  For this purpose we   study here in details the spectra  of the corresponding transfer operators.  Finally, in section~\ref{Sec8} we give the concluding remarks.  

\section{The kicked quantum map on a lattice}\label{Sec3}
To start our consideration we introduce a multidimensional lattice model of periodically kicked locally interacting particles. The system  Hamiltonian, $H(t)$,  consists of the time-dependent and the time-independent parts,
\begin{equation}\label{Ham}
H(t)=H_I+H_K\sum_{\tau=-\infty}^{+\infty}\delta(t-\tau).
\end{equation}
The kick part of the Hamiltonian, $H_K$, induces independent evolution of non-interacting particles. It turns on periodically at integer instants of time $\tau$. The $H_I$ part (eq.~\ref{Ham}) describes the nearest neighbour interaction between the particles. Such Hamiltonian structure, in particular, implies that the quantum time evolution can be written as a product,
\begin{equation}\label{Unitgeneric}
U =U _KU_I,
\end{equation}
of unitary evolutions~\cite{AWGG16, AWGBG16, AGBWG18} $U_K$ and $U_I$, corresponding to the kick and the interaction parts of the Hamiltonian, respectively. 

To specify the form of the unitary operators we define, first, the on-site local Hilbert space $\mathcal{H}$ equipped with the discrete $L$-dimensional basis $\set{\ket{s},\; s=\overline{1,L}}$. The total Hilbert space of the system is defined then by the tensor product $\H^{\otimes \N}$. It has the dimension $L^\mathcal{N}$ and possesses the natural product basis,
\begin{equation} \set{\ket{\bm s}\equiv\prod_{\bm j\in\Lat} \ket{s_{\bm j}} , \,\,\,  s_j=\overline{1,L}}, \label{basis}\end{equation}
 where the multidimensional index $\bm j$ marks the particles' positions in the lattice $\Lat$ and the product runs over all $\N$ lattice sites.  The Floquet time evolution between the kicks is governed by the unitary operator $U_I=e^{-\i H_I}$. We require that $H_I$ couples the nearest neighbour sites of the multidimensional lattice and has to be diagonal in the product basis (\ref{basis}). This yields the following matrix form of the evolution operator:
\begin{multline}\label{UIFloquet}
\braket{\bm s|U_I[\bm f]|\bm s'}= \\=\delta(\bm s,\bm s')\prod_{d=1}^D \exp\left[\i  \sum_{\bm j}f_d(s_{\bm j},s_{\bm j+\bm 1_d})\right],
\end{multline}
where $U_I[\bm f]$ is determined by the set of   functions $\bm f= (f_1,f_2, \dots, f_D)$. Here we used the notation $\bm 1_d$, which denotes the one site shift of the index $\bm j$ in the spatial direction $d$. The function $\delta(\bm s,\bm s')$ stands for the product of the Kronecker symbols, $\delta(\bm s,\bm s')=\prod_{\bm j}\delta(s_{\bm j},s'_{\bm j})$. In the above formula (eq.~\ref{UIFloquet}) the cyclic boundary conditions in each spatial dimension are implied. 

Note that, to satisfy the unitary condition for $U_I$, each function $f_d$ in eq.~(\ref{UIFloquet}) has to be a real-valued function. The operator $U_I$ acts independently in each spatial direction so that it can be represented as a product of mutually commuting unitary operators $U_{I_d}$, where $d$ indexes the spatial axis number, 
\begin{multline}
U_I[\bm f]=\prod_{d=1}^D U_{I_d},\\
\braket{\bm s| U_{I_d}|\bm s'}=\delta(\bm s,\bm s')\exp\left[\i  \sum_{\bm j}f_d(s_{\bm j},s_{\bm j+\bm 1_d})\right].\label{Uinter}
\end{multline}

The kick part, $H_K$, of the total Hamiltonian $H$ in eq.~(\ref{Ham}) defines the on-site particle dynamics. The corresponding evolution operator, $U_K=e^{-\i H_K}$, can be represented as the tensor product of unitary operators defined in the single-particle Hilbert space. We assume that the kicks act identically on each particle, so that
\begin{equation}\label{UKprod}
\braket{\bm s|U_K[g]|\bm s'}=\prod_{\bm j}\braket{s_{\bm j}|u[g]|s'_{\bm j}},
\end{equation}
where $u$ is the $L\times L$ unitary matrix with the symbol $g$. Its entries, 
\begin{equation}\label{usmall}
\braket{s|u[g]|s'}=\frac{1}{\sqrt{L}}e^{\i g(s,s')},
\end{equation}
satisfy thy unitarity condition
\begin{equation}\label{usmall2}
\frac{1}{L}\sum_{s'=1}^L e^{\i g(s,s')}e^{-\i g^*(s'',s')}=\delta(s,s'').
\end{equation}

Having introduced the basic notations we are in a position to formulate the duality relation for the quantum maps, which has been established for the one-dimensional case in ref.~\cite{GBAWG20}. In the case  $D=1$ it states that for the dual-unitary quantum system the evolution operator $U=U_K[g]U_I[f]$  and its dual counterpart, $\tilde U=U_K[f]U_I[g]$, obtained by the exchange of $f$ and $g$ functions, are both unitary. Note that this requires the Hadamard property for the matrices 
\begin{equation}
    \bra{s}u[f]\ket{s'}=\frac{e^{\i f(s,s')}}{\sqrt{L}}, \quad 
\bra{s}u[g]\ket{s'}=\frac{e^{\i g(s,s')}}{\sqrt{L}}.
\end{equation}
In other words, these are $L\times L$ unitary matrices with identical absolute values for all entries, where both $f$ and $g$ are real-valued functions. If, furthermore, $f=g$ we have $U=\tilde U$ and the system possesses the  spatiotemporal symmetry. 
  
The above notion of dual-unitarity can be straightforwardly extended to an arbitrary dimension $D$. In general, there are exist $D$ possible dual operators,
\begin{equation}
    \tilde U_j=U_K[ f_j]U_I[\bm {\tilde g}], \qquad \bm {\tilde g}=(f_1,\dots, g,\dots, f_D),
\end{equation}
where the function $g$ is exchanged with one of the spatial coupling functions $f_j$. We say that the system is partially dual unitary if one of the spatial evolution operators $\tilde U_j$ is unitary. The system is called fully dual-unitary if all $\tilde U_j, j=1,\dots, D$  are unitary. If, in addition  all functions $f_j, j=1,\dots,D$ and $g$ are equal, then
\[U=\tilde U_1=\tilde U_2= \dots =\tilde U_D.\]
In this case, the system possesses full spatiotemporal symmetry.

\section{Correlations between local operators}\label{Sec4}
In this paper, we  aim at the calculation of the correlation function, 
\begin{equation}\label{CorrFunc}
C(t)=L^{-\mathcal{N}}\:\tr \bar{\Sigma} U^{-t} \ubar{\Sigma}U^t,
\end{equation}
for two local observables $\bar{\Sigma},\ubar{\Sigma}$. For simplicity of exposition, we restrict our consideration to the case of  two-dimensional spatial lattices $\bar{\mathbb{Z}}^2$, while the generalization to $D$-dimensional lattices with $D>2$ is straightforward. 

\subsection{One-dimensional lattice}
We start by recalling the results of \cite{GBAWG20}  for the one-dimensional case.
 There the correlation function (\ref{CorrFunc}) was calculated for the operators  $\bar{\Sigma}$ and $\underline{\Sigma}$ given by the products of the following form
\begin{eqnarray}\label{bSigma}
\bar{\Sigma}&=&\q_1\otimes \q_2\otimes \underbrace{\1\otimes\dots\otimes\1}_{N-2};\\
\ubar{\Sigma}&=&\underbrace{\1\otimes\dots\otimes\1}_{n}\otimes \q_3\otimes \q_4\otimes \underbrace{\1\otimes\dots\otimes\1}_{N-n-2},\label{Sigmab}
\end{eqnarray}
where each $\q_\ell$ ($\ell=1,2,3,4$) is an operator acting on the on-site Hilbert space $\H$. 

For the one-dimensional ($D=1$) dual unitary model the correlation function (\ref{CorrFunc}) for the traceless $q_\ell$  always equals zero, except for the case when the correlations are considered along the ``light-cone'' edge. The latter case corresponds to the choice  $\abs{n}=t$  in eq.~(\ref{Sigmab}).  The resulting   correlation function  at the light-cone edge, $n=t>2$, can be represented as the expectation value of the transfer operator, $\bm T$, power
\begin{equation}\label{CD1}
C_{D=1}(t)=\bra{\bar{\bm\Phi}_{\q_1,\q_2}}\bm T^{t-2}\ket{\bm \Phi_{\q_3,\q_4}}, 
\end{equation}
where the vectors  $\bra{\bar{\bm\Phi}_{\q_1,\q_2}},\ket{\bm \Phi_{\q_3,\q_4}} $ depend on the operators $\q_1$, $\q_2$, and $\q_3$, $\q_4$, respectively.
The transfer operator $\bm T$ acts in the Hilbert space $\mathcal{H}\otimes \mathcal{H}$, and has the entries
\begin{multline}\label{TD1}
\bra{\chi,\eta}\bm T\ket{\chi',\eta'}\\=\frac{1}{L^3}\abs{\sum_{s=1}^L e^{\i f_1(\chi,s)+\i g(\eta,s)+\i g(s,\chi')+\i f_1(s,\eta')}}^2.
\end{multline}

 In a general non-dual case the correlation function becomes zero only outside the light-cone, $\abs{n}>t$  (which is supported by the argument that the speed of information propagation in the kicked chain model equals 1), but remains finite   inside the light cone, $\abs{n}<t$.

The result (\ref{CD1}) was obtained in~\cite{GBAWG20} by representing the correlation function $C(t)$ in the form of a partition function of a classical spin lattice model, followed by the application of contraction rules that allow the elimination of most of the spin variables.
As we demonstrate below this method   also applies to  the  higher dimensional lattices.

\subsection{Two-dimensional lattice} We now consider the correlation function  (\ref{CorrFunc}) for two-dimensional lattice models.  For convenience we address two spatial directions as ``vertical'' and ``horizontal'', using indexes $v$ and $h$, instead of 1 and 2. The number of sites in the vertical and horizontal direction is denoted by $N$ and $M$, respectively, with $\N=MN$ being the total number of lattice sites. The two-point correlation function is defined by  eq.~(\ref{CorrFunc}) with the evolution matrix $U=U_KU_I$, where  the interaction part of the evolution operator, 
\begin{multline}\label{UI2D}
\bra{\bm s}U_I\ket{\bm s'}\equiv \delta(\bm s,\bm s')\\ \times \prod_{n=1}^N\prod_{m=1}^M e^{\i f_v(s_{n,m},s_{n+1,m})+ f_h(s_{n,m},s_{n,m+1})},
\end{multline}
is diagonal in the product basis $\ket{\bm s}$.
The circular boundary conditions are assumed in the above formulae, namely, $s_{n,M+1}\equiv s_{n,1}$, and $s_{N+1,m}\equiv s_{1,m}$.
The kicked part of the evolution is defined by the eqs.~(\ref{UKprod}),~(\ref{usmall}).

We consider correlations (\ref{CorrFunc}) for a pair of  many-body operators $\bar{\Sigma}$ and $\underline{\Sigma}$, each  supported at four neighbouring  points of the lattice. They are   defined by  eight (rather than  four, as in eqs.~\ref{bSigma},~\ref{Sigmab}) local  matrices $\q_\ell$, $\ell=1,\dots,8$. Without loss of generality, we define $\bar{\Sigma}$ as a direct product, where the non-trivial matrices are placed in the ``left upper corner'', i.e. at the coordinates $(n,m)=\set{(1,1),(1,2),(2,1),(2,2)}$. The second many-body operator, $\ubar{\Sigma}$ has the non-trivial entries at the coordinates $(n,m)=\set{(\nu,\mu),(\nu,\mu+1),(\nu+1,\mu),(\nu+1,\mu+1)}$. Schematically, these operators can be represented as follows
\begin{widetext}
\begin{equation}\label{barSigma}
\begin{array}{l}
\\
\bar{\Sigma}=\end{array}\begin{array}{cccccccccc|r}
1&&2&&3&&&&N&\\\hline
\q_1&\otimes&\q_2&\otimes&\1&\otimes&\dots&\otimes&\1&\otimes&\;1\\
\q_3&\otimes&\q_4&\otimes&\1&\otimes&\dots&\otimes&\1&\otimes&\;2\\
\1&\otimes&\1&\otimes&\1&\otimes&\dots&\otimes&\1&\otimes&\;3\\
&&&&&&\vdots&&&&\\
\1&\otimes&\1&\otimes&\1&\otimes&\dots&\otimes&\1&&\;M\\
\end{array},\qquad
\begin{array}{l}
\\
\underline{\Sigma}=\end{array}\begin{array}{ccccccccccc|r}
&&\nu-1&&\nu&&\nu+1&&\nu+2&&\\
\hline
&&\vdots&&\vdots&&\vdots&&\vdots&&&\\
\dots&\otimes&\1&\otimes&\1&\otimes&\1&\otimes&\1&\otimes&\dots&\;\mu-1\\
\dots&\otimes&\1&\otimes&\q_5&\otimes&\q_6&\otimes&\1&\otimes&\dots&\;\mu\\
\dots&\otimes&\1&\otimes&\q_7&\otimes&\q_8&\otimes&\1&\otimes&\dots&\;\mu+1 \\
\dots&\otimes&\1&\otimes&\1&\otimes&\1&\otimes&\1&\otimes&\dots&\;\mu+2 \\
&&\vdots&&\vdots&&\vdots&&\vdots&&&
\end{array} 
\end{equation}
 \end{widetext}
 
To evaluate the correlation function, we, first,  introduce the basis vectors additionally indexed by the integer time  $t$, 
\begin{equation} \ket{\bm{s}_t}\equiv\prod_{(n,m)\in\Lattwo} \ket{\bm{s}_{nmt}}, \label{2Dbasis}\end{equation}
and then write
\begin{multline}\label{tracedefinition}
C(T)\equiv L^{-\N}\tr\bar{\Sigma} U^{-T} \ubar{\Sigma}U^T
\\=L^{-\N}\sum\bra{\bm s_{2T+1}}\bar{\Sigma} \ket{\bm s_0}  \bra{\bm s_{T}} \ubar{\Sigma} \ket{\bm s_{T+1}}\\\times\left(
\prod_{t=0}^{T-1} \bra{\bm s_t}U^\dag\ket{\bm s_{t+1}}\bra{\bm s_{2T-t}} U\ket{\bm s_{2T-t+1}}\right),
\end{multline}
where the sum runs over all possible values of the components, $s_{nmt}\in \overline{1,L}$. We assume that for $T=0$ the product in eq.~(\ref{tracedefinition}) equals $1$ identically. It is also convenient to introduce at each position $(m,n)$ a two-component spin variable with the components $\bar{s}_{nmt}$ and $\ubar{s}_{nmt}$. The upper component of the spin variable, $\bar{s}_{n,m,t}$ coincides with $s_{n,m,t}$ for all $t\in\overline{0,T}$. The lower component  corresponds to the spin variable  taken at the conjugated instance of time $2T-t$, such that $\ubar{s}_{n,m,t}=s_{n,m,2T-t+1}$ with the same set of the indexes  $t\in\overline{0,T}$. The schematics of the above correspondence are plotted below for convenience
{\small \begin{equation}
\begin{array}{ccccccccccccc}
\bm s_0&&&\bm s_{T-1}&&\bm s_{T} &\; |\;&\bm s_{T+1}& &\bm s_{T+2} & & &\bm s_{2T+1}\\
\bullet&\Leftrightarrow&\dots&\bullet&\Leftrightarrow&\bullet&\; |\;&\bullet&\Leftrightarrow&\bullet&\dots&\Leftrightarrow&\bullet \\
\bar{\bm s}_0& & &\bar{\bm s}_{T-1}& &\bar{\bm s}_{T} &\; |\;&\ubar{\bm s}_{T}& &\ubar{\bm s}_{T-1} & & &\ubar{\bm s}_0 
\end{array}
\end{equation}}
Using this notation  we can rewrite the correlation function in a symmetric form,
\begin{multline}\label{tracesum1}
C(T)=L^{-\N}\sum_{ \Sp}\bar{\Phi}(\bar{\bm s}_0,\ubar{\bm s}_0)\bar{T}_I(\bar{\bm s}_{0},\ubar{\bm s}_0)\Phi(\bar{\bm s}_{T},\ubar{\bm s}_{T} )\\\times \prod_{t=0}^{T-1}\bar{T}_K(\bar{\bm s}_{t},\bar{\bm s}_{t+1};\ubar{\bm s}_{t+1},\ubar{\bm s}_{t})\bar{T}_I(\bar{\bm s}_{t+1},\ubar{\bm s}_{t+1}) ,
\end{multline}
where $\bar{\Phi}$ and $\Phi$ depend on $\N$ spin variables and implicitly include dependence on the matrices $q_\ell$, 
\begin{eqnarray}\label{Phifunctions}
\bar{\Phi}(\ubar{\bm s}_0,\bar{\bm s}_0)&=&\bra{\ubar{\bm s}_{0}}\bar{\Sigma} \ket{\bar{\bm s}_0};\\
\Phi(\bar{\bm s}_{T},\ubar{\bm s}_{T})&=&\bra{\bar{\bm s}_{T}}U_I \ubar{\Sigma} U_I^\dag\ket{\ubar{\bm s}_{T}}.
\end{eqnarray}
Each $\bar{T}_I$ is a function of  $\N$ spin variables and describes the particle interactions between the kicks. Substituting the explicit form of the matrix $U_I$ entries  (eq.~\ref{UI2D}) we have for $\bar{T}_I$ \begin{multline}\label{TIbarExplicit}
\bar{T}_I(\bar{\bm s}_{t},\ubar{\bm s}_t)\equiv 
 \bra{\bar{\bm s}_t}U_I \ket{\bar{\bm s}_{t}}^*\bra{\ubar{\bm s}_{t}} U_I\ket{\ubar{\bm s}_{t}} 
\\=
\prod_{n=1}^{N}\prod_{m=1}^{M}e^{- \i f_v(\bar{s}_{n,m,t},\bar{s}_{n+1,m,t})+\i f_v(\ubar{s}_{n,m,t},\ubar{s}_{n+1,m,t})}\\\times e^{-\i f_h(\bar{s}_{n,m,t},\bar{s}_{n,m+1,t})+\i  f_h(\ubar{s}_{n,m,t},\ubar{s}_{n,m+1,t})}.
\end{multline}
The explicit form of the matrix $U_K$ (eqs.~\ref{UKprod},~\ref{usmall}) allows to write down the expression for $\bar{T}_K$:
\begin{multline}\label{TbarExplicit}
\bar{T}_K(\bar{\bm s}_{t},\bar{\bm s}_{t+1};\ubar{\bm s}_{t+1},\ubar{\bm s}_t)\equiv   \bra{\bar{\bm s}_{t+1}}U_K \ket{\bar{\bm s}_{t}}^* \bra{\ubar{\bm s}_{t+1}} U_K\ket{\ubar{\bm s}_{t}}  
\\= 
\frac{1}{L^{NM}}\prod_{n=1}^{N}\prod_{m=1}^{M} e^{- \i  g^*(\bar{s}_{n,m,t+1},\bar{s}_{n,m,t}) +\i  g(\ubar{s}_{n,m,t+1},\ubar{s}_{n,m,t})}.
\end{multline}
The sum runs over all possible values of the full  set 
\begin{equation}
    \Sp=\set{(\bar{s}_{nmt}, \ubar{s}_{nmt})| (n,m,t) \in \mathcal{L} },
\end{equation}
of the $MNT\equiv|\Sp|$ spin variables,
located at the  nodes of the 3D space-time grid:
\begin{equation}
\mathcal{L}=\{(n,m,t)| t\in \overline{0, T}, n\in\overline{1,N}, m\in\overline{1,M}\}.
\end{equation} 
Note, that in the present calculations, we have used the symmetrization procedure slightly different than in Ref.~\cite{GBAWG20}. Namely, instead of incorporating $U_I$ into $\bar{\Phi}$ we introduced an additional unity operator $\1=U_IU_I^\dag=U_I^\dag U_I$ from both sides of $\ubar{\Sigma}$. This is done to formulate the contraction rules in a symmetric way.

The correlation function (\ref{tracesum1}) can be equally represented in another form with the structure of a partition function, 
\begin{equation}
    C(T)=\frac{1}{L^{|\Sp|}}
    \sum_{\Sp}G_1(\Sp_1) G_2(\Sp_2) e^{-i\mathcal{F}(\Sp)}.\label{method1body}
\end{equation}
The expression under the sum is split into the product of three factors in accordance with the location of the spin variables within the lattice $\mathcal{L}$.  
The first one  is given by, $G_1(\Sp_1) =D_{\bar{\Sigma}} D_{\ubar{\Sigma}}$, where
\begin{multline}
D_{\bar{\Sigma}}= \langle \bar{s}_{111}| u[g] \q_1u^\dag[g] |s_{111}\rangle \langle \bar{s}_{121}|u[g]  \q_2u^\dag[g] |s_{121}\rangle\\
\langle \bar{s}_{211}|u[g]  \q_3u^\dag[g] |s_{211}\rangle \langle \bar{s}_{221}| u[g] \q_4u^\dag[g] |s_{221}\rangle, 
\end{multline} 
\begin{multline}
D_{\ubar{\Sigma}}= \langle \bar{s}_{\nu\mu T}| \q_5 |s_{\nu\mu T}\rangle \langle \bar{s}_{\nu\mu+1 T}| \q_6 |s_{\nu\mu+1 T}\rangle\\
\langle \bar{s}_{\nu+1\mu T}| \q_7 |s_{\nu+1\mu T}\rangle \langle \bar{s}_{\nu+1\mu+1 T}| \q_8 |s_{\nu+1\mu+1 T}\rangle, 
\end{multline} 
It depends on the spin variables,
\begin{equation}
    \Sp_1=\set{(\bar{s}_{nmt}, \ubar{s}_{nmt})| (n,m,t) \in \mathcal{L}_1 },
\end{equation}
located at the eight   lattice sites, 
\begin{multline} \mathcal{L}_1=\{(1,1,1),(1,2,1),(2,1,1),(2,2,1), (\nu,\mu,T),\\(\nu,\mu+1,T), (\nu+1,\mu,T),(\nu+1,\mu+1,T)\},\nonumber\end{multline} 
corresponding to the position of the local operators $\q_\ell$ in $\bar{\Sigma}$, $\ubar{\Sigma}$.  The second term is given by the  product of the Kronecker delta functions
 \begin{equation}
     G_2(\Sp_2)= \prod_{(n,m,t)\in\mathcal{L}_2}\delta(\ubar{s}_{nmt},\bar{s}_{nmt}).
 \end{equation}
It depends on the spin variables 
\begin{equation}
    \Sp_2=\set{(\bar{s}_{nmt}, \ubar{s}_{nmt})| (n,m,t) \in \mathcal{L}_2 },
\end{equation}
 located at  the  subset 
   \[\mathcal{L}_2=\{(n,m,t)| t\in\{1,T\}, n\in\overline{1,N}, m\in\overline{1,M}\} \setminus \mathcal{L}_1\]
   of the 3-dimensional space-time grid.
 Finally,  the interaction between spins is described by the third  term $e^{-i\mathcal{F}(\Sp)}$, where    the exponent is given by the  function  
\begin{multline}\mathcal{F}=\sum_{(n,m,t)\in\mathcal{L}}
f_v(\bar{s}_{n,m,t},\bar{s}_{n+1,m,t})+ f_v(\ubar{s}_{n,m,t},\ubar{s}_{n+1,m,t})\\+f_h(\bar{s}_{n,m,t},\bar{s}_{n,m+1,t})+  f_h(\ubar{s}_{n,m,t},\ubar{s}_{n,m+1,t})\\
+g^*(\bar{s}_{n,m,t+1},\bar{s}_{n,m,t}) +  g(\ubar{s}_{n,m,t+1},\ubar{s}_{n,m,t})
,\label{expfunction}
\end{multline}
 and the sum runs over the full set $\Sp$ of spin variables.

 \section{Graphical method for evaluation of correlations}\label{Sec5}

\begin{figure}
\begin{center}
    \includegraphics[scale=0.38]{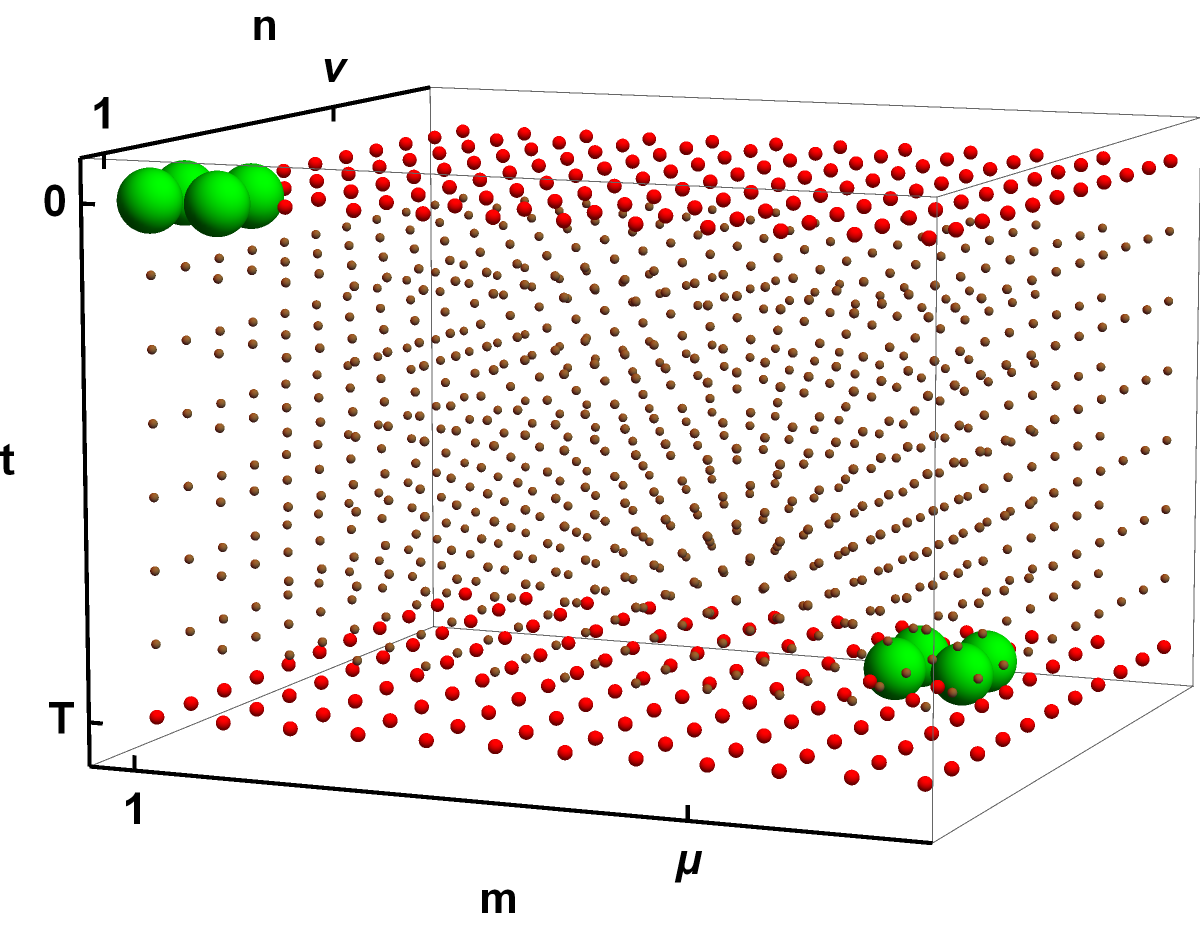}
\end{center}
\caption{\label{Fig1}\small 3D grid of the configuration space for calculation of the sum in eq.~(\ref{tracesum1}) for the choice of $N=11$, $M=12$, $T=7$, $\nu=8$, and $\mu=9$. The brown small balls correspond to the uncorrelated spin components, the red balls show the presence of the Kronecker symbol $\delta(\bar{s}_{n,m,t},\underline{s}_{n,m,t})$ at the corresponding vertex $(n,m,t)$. The large green balls denote the positions of the matrices $q_\ell$ entering the many-body operators $\bar{\Sigma}$ at $t=T$, and $\ubar{\Sigma}$ at $t=0$ (eqs.~\ref{barSigma}). The function $\bar{T}_I$ (eq.~\ref{TIbarExplicit}) defines the amplitudes of interaction between the balls in each time-plane, while $\bar{T}_K$ (eq.~\ref{TbarExplicit}) is responsible for interaction between the time-planes. }
\end{figure}

The correlation function, expressed   in the form of the partition function (\ref{method1body}) permits  an instructive graphical representation, as illustrated in fig.~\ref{Fig1}.  The  3D space-time  lattice  $\mathcal{L}$ is partitioned  into  the three subsets $\mathcal{L}_1, \mathcal{L}_2, \mathcal{L}_3 \equiv \mathcal{L}/(\mathcal{L}_1\cup\mathcal{L}_2)$. In the picture, the spin variables are schematically shown by balls of various colours in accordance with their role in the partition function~(\ref{method1body}).  The green-coloured balls, located at $\mathcal{L}_1$, correspond to the positions of the operators $q_\ell$, see eq.~\ref{barSigma}. The red balls, located at the subset $\mathcal{L}_2$, correspond to the $\delta$-correlated spins. Since each unit matrix $\1$ in the operators $\bar{\Sigma}$ and $\underline{\Sigma}$ generates the pair of the $\delta$-correlated spins, the majority of the endpoints are coloured red. Finally, the brown balls placed at $\mathcal{L}_3$ correspond to all other spin variables. 

In general, the partition function~(\ref{method1body}) can be calculated by eliminating the spin variables one by one. To facilitate this procedure, a simple graphical method  can be developed by drawing an analogy with Ref.~\cite{GBAWG20}. Below, we formulate the ``contraction rules", which form the basis of our approach.

\subsection{ Contraction rules} 
\begin{figure*} 
\begin{tabular}{lc}
a)&\includegraphics[scale=0.80]{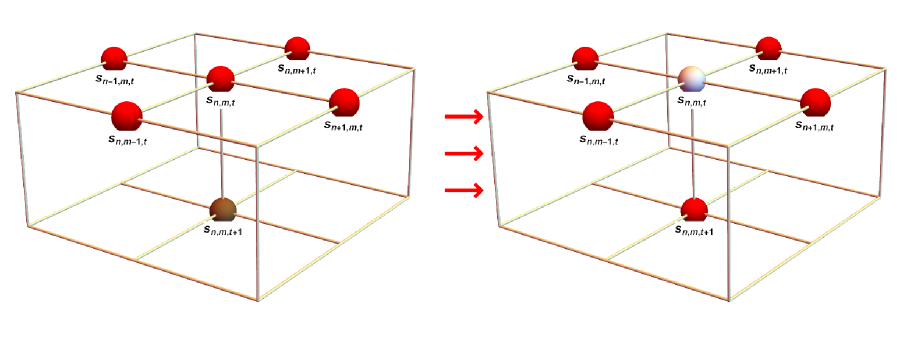}  \\
b)&\includegraphics[scale=0.80]{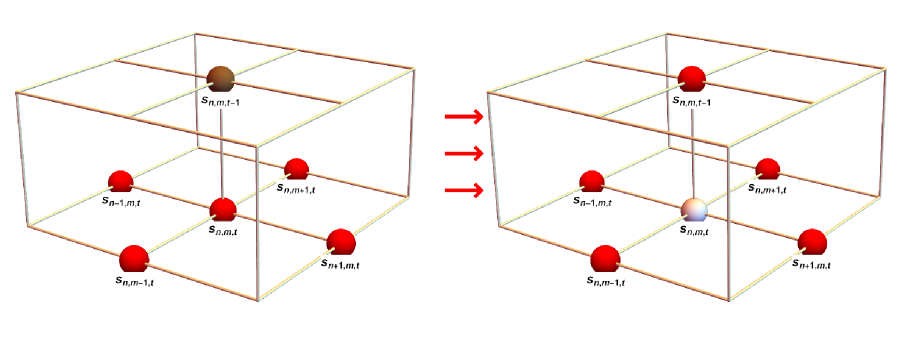}  \\
c)&\includegraphics[scale=0.80]{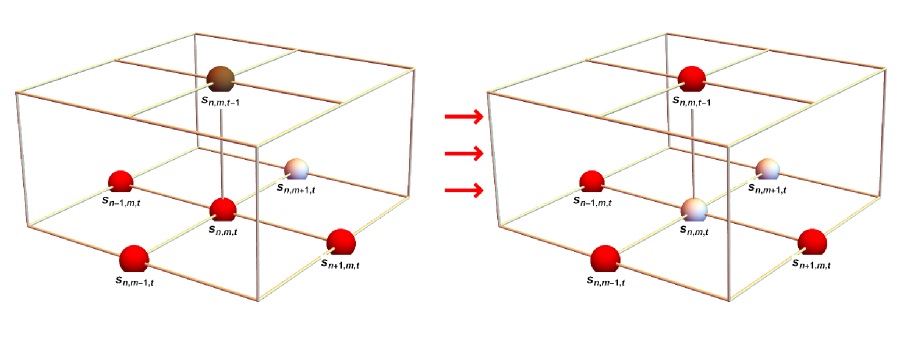} 
\end{tabular}
\caption{\label{FigGrid}\small The graphical representation of the contraction rules in the time-direction. The summation over the correlated spins $\bar{s}_{n,m,t}$ and $\ubar{s}_{n,m,t}$  (the central red balls in the plots from the left column) surrounded by other correlated spins transforms the initially uncorrelated spins (brown balls of the plots in the left column) into the delta-correlated ones. The white balls in the final configurations (right column of the plots) show the spins over which the summation has been performed. It is assumed that summation over the spins on the previous level (the spins $\bar{s}_{n,m,t-1}$, $\ubar{s}_{n,m,t-1}$ for the diagram (a) and the spins $\bar{s}_{n,m,t+1}$, $\ubar{s}_{n,m,t+1}$ for the diagrams (b,c)) has been done to generate a white or red ball at the corresponding position.}
\end{figure*}

Consider the local configuration consisting of four neighbouring $\delta$-correlated spins and an unpaired spin $(\bar{s}_{n,m,t+1},\underline{s}_{n,m,t+1})$, as it is shown in fig.~\ref{FigGrid}a.  The corresponding part of the partition function   includes five $\delta$-symbols, eight phases entering the function $\bar{T}_I(\bar{\bm s}_t,\ubar{\bm s}_t)$ and two phases from the function $\bar{T}_K(\bar{\bm s}_{t},\bar{\bm s}_{t+1};\ubar{\bm s}_{t+1},\ubar{\bm s}_{t})$: 
\begin{multline}\label{contractionrule1}
\Gamma=L^{-1}\sum_{s=1}^L
e^{- \i \big( g^*(\bar{s}_{n,m,t+1},s)- g(\ubar{s}_{n,m,t+1},s)\big)}\\
\times\delta(\bar{s}_{n+1,m,t},\underline{s}_{n+1,m,t})\delta(\bar{s}_{n,m+1,t},\underline{s}_{n,m+1,t})\\\times \delta(\bar{s}_{n-1,m,t},\underline{s}_{n-1,m,t})\delta(\bar{s}_{n,m-1,t},\underline{s}_{n,m-1,t})
\\\times
e^{- \i\big( f_v(s,\bar{s}_{n+1,m,t})- f_v(s,\ubar{s}_{n+1,m,t})\big)}\\\times  e^{-\i\big( f_v(\bar{s}_{n-1,m,t},s)- f_v(\ubar{s}_{n-1,m,t},s)\big)}\\\times e^{-\i \big( f_h(s,\bar{s}_{n,m+1,t})- f_h(s,\ubar{s}_{n,m+1,t})\big)}\\\times  e^{- \i \big( f_h(\bar{s}_{n,m-1,t},s)- f_h(\ubar{s}_{n,m-1,t},s)\big)}.
\end{multline}
Utilizing the unitarity of the kick evolution operator $u[g]$  (see eq.~\ref{usmall2}) and the properties of the Kronecker function the sum in eq.~(\ref{contractionrule1}) can be simplified to:
\begin{equation}
 \Gamma = \delta(\bar{s}_{n,m,t+1},\underline{s}_{n,m,t+1}).   \label{contractionrule}
\end{equation}
Graphically, this contraction rule can be interpreted as a transition from the configuration shown in fig.~\ref{FigGrid}a (left),  to the configuration depicted in fig.~\ref{FigGrid}a (right), where the spin variable $(\bar{s}_{n,m,t+1},\underline{s}_{n,m,t+1})$ has been eliminated. By analogy, the contraction rule can be formulated in the opposite time direction ($t\to t-1$), see fig.~\ref{FigGrid}b.

Note also that similar contraction rules exist for the configurations with a reduced number of spin variables, see fig.~\ref{FigGrid}c. In fact, a white ball (where a spin variable was eliminated in the previous steps) at some node $(n,m,t)$ can be replaced back by the identity multiplier presented as a sum of the Kronecker symbols, $1\equiv L^{-1}\sum_{\bar{s}_{n,m,t},\underline{s}_{n,m,t}=1}^L\delta(\bar{s}_{n,m,t},\underline{s}_{n,m,t})$. This allows to extend the contraction rule for all combinations of balls including those, where some red balls are substituted by the white ones. 

\subsection{Application of the contraction rules} 
\begin{figure*}
\begin{tabular}{lccc}
a)&\includegraphics[scale=0.32]{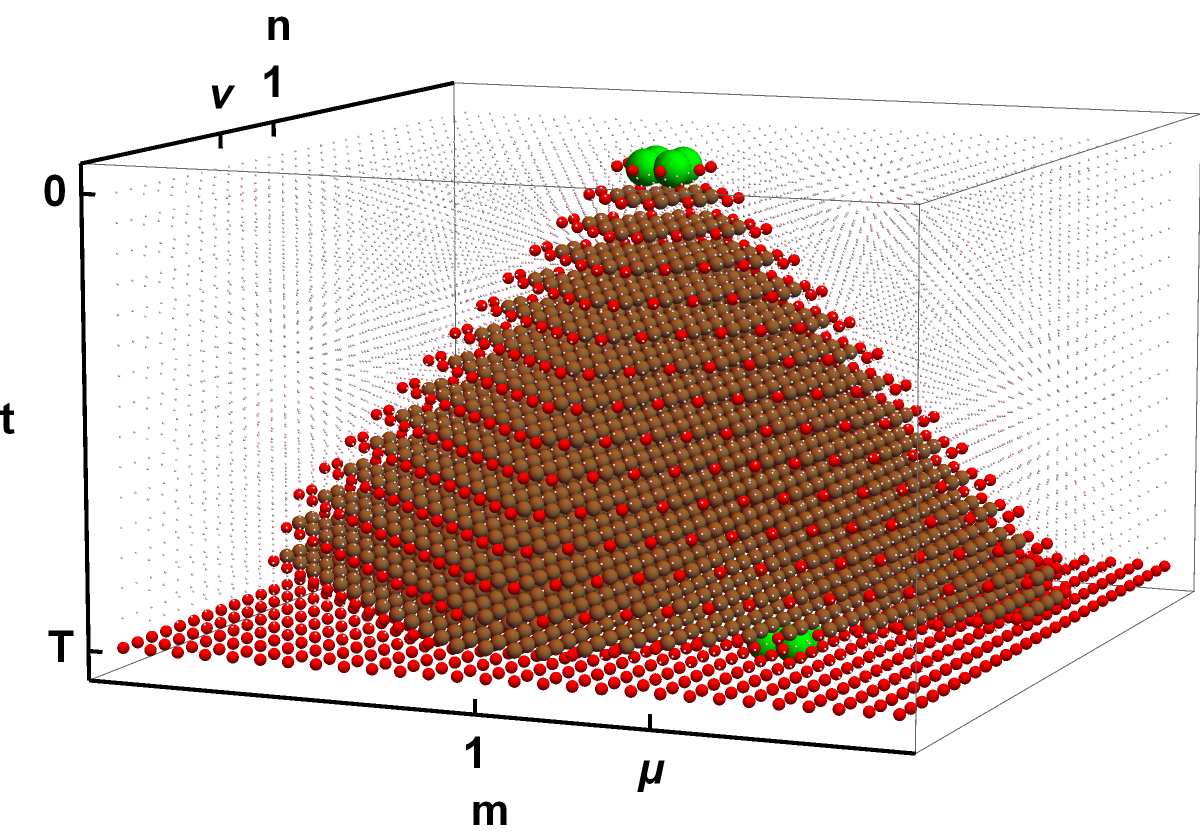}&b)&\includegraphics[scale=0.32]{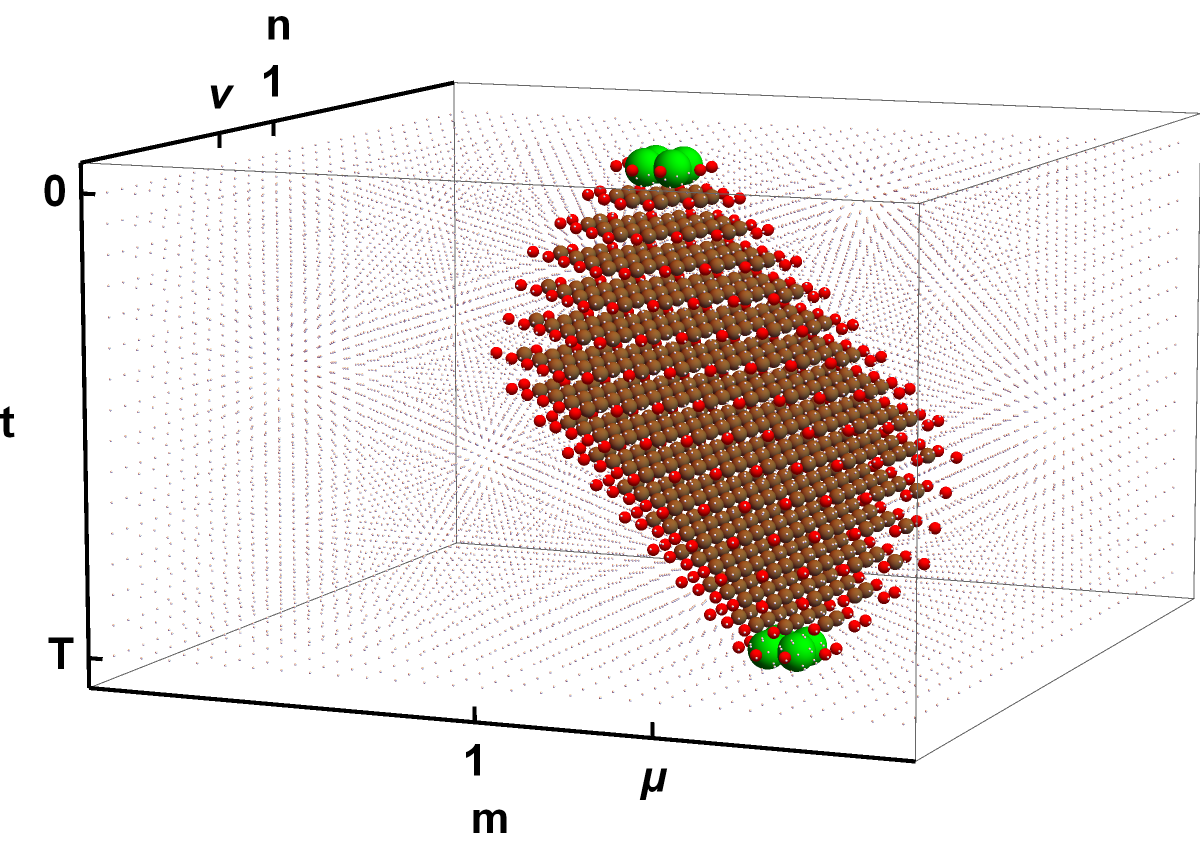}
\end{tabular}
\caption{\label{Fig2}\small a) The spin configuration obtained after consequent application of the contraction rule (fig.~\ref{FigGrid}a) to the initial configuration (fig.~\ref{Fig1}) with the parameters $N=M=28$, $T=15$, $\nu=25$, and $\mu=7$. The small white points have the same meaning as in the fig.~\ref{FigGrid}. For better representation, the spin array was periodically shifted along the horizontal and vertical axes. b) The spin configuration after the iterative application of the contraction rule (fig.~\ref{FigGrid} b, c) to the structure given in the plot a.}
\end{figure*}

We now proceed with eliminating the spin variables from the partition function (\ref{method1body}) by using the contraction rules graphically formulated in figs.~\ref{FigGrid}.  
Application of the contraction rules along the time axis, starting from $t=0$ and continuing up to $t=T-1$ allows us to eliminate most of the spin variables, resulting in the pyramidal structure shown in fig.~\ref{Fig2}a. The coordinates $(n,m,t)$ of the remaining spin variables satisfy the equation $t+1\ge \abs{n-3/2}+\abs{m-3/2}$ (the positions of $q_\ell$ are fixed by eqs.~\ref{barSigma}). Furthermore,  the application of the contraction rule in the opposite time direction, i.e. $t$ changes from $T$ to $0$, yields a parallelepiped-like structure shown in fig.~\ref{Fig2}b.  This parallelepiped shrinks to a line when the operators $\q_\ell$ ($\ell=5,6,7,8$) are positioned at one of the vertices of the pyramid base. If the operators $\q_\ell$ ($\ell=5,6,7,8$) are placed outside the pyramid base, all spin variables can be eliminated by contraction rules and the resulting correlation function nullifies (for the traceless $\q_\ell$). Note that this observation is consistent with the causality argument given in Sec.~\ref{main_ideas}.

\subsection{Dual-unitary case}
\begin{figure*} 
\begin{tabular}{cccc}
a)&\includegraphics[scale=0.45]{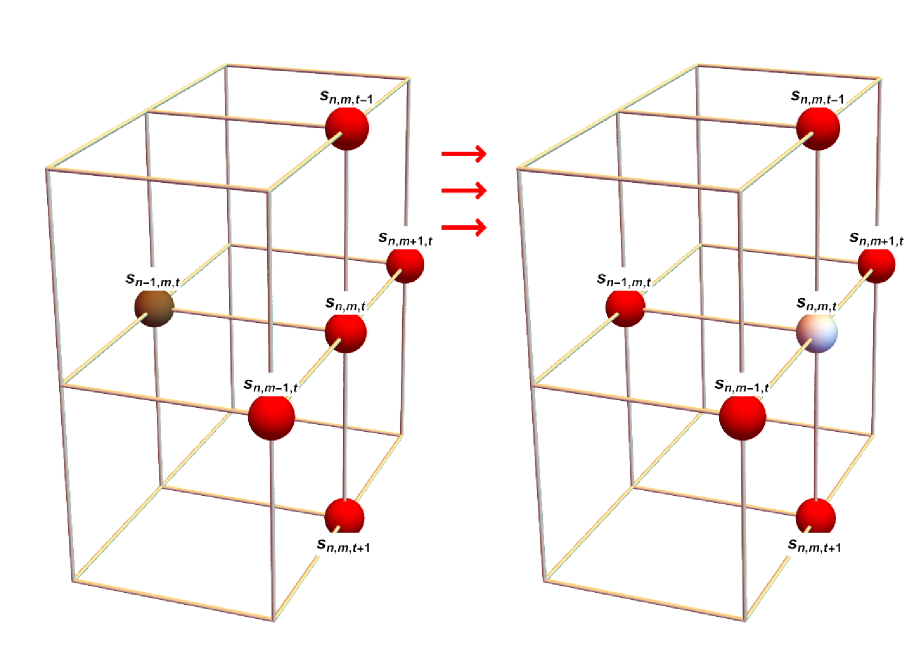} & b)&\includegraphics[scale=0.45]{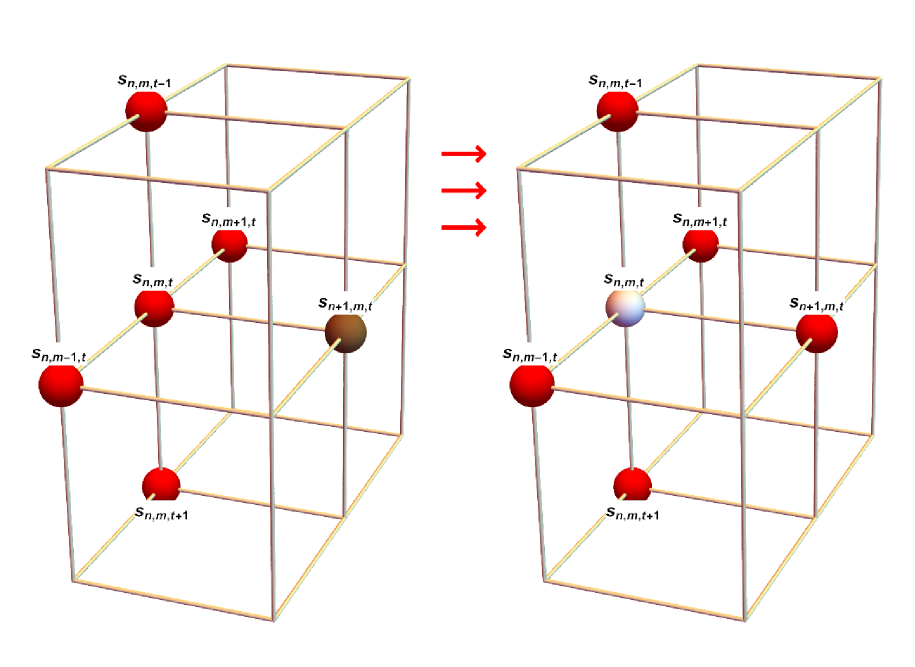} 
\end{tabular}
\caption{\label{FigGrid1}\small The graphical representation of the contraction rules in the spatial direction $n$, dual to the time direction. The summation over the correlated spins $\bar{s}_{n,m,t}$ and $\ubar{s}_{n,m,t}$  (the central red balls in the plots from the left column) transforms the initially uncorrelated spins (brown balls) into the correlated ones. The white balls in the final configurations (the right column of the plots) show the spins over which the summation has been performed. It is assumed that summation over the spins on the previous level (the spins $\bar{s}_{n-1,m,t}$ and $\ubar{s}_{n-1,m,t}$ for the diagrams a, b and the spins $\bar{s}_{n+1,m,t}$ and $\ubar{s}_{n+1,m,t}$ for the diagrams c, d) has been already done to generate a white/red ball at the corresponding position.}
\end{figure*}

Up to now, we considered a general case, where no further systematic elimination of the spin variables can be done in the  sum (\ref{method1body}). Further progress in calculations becomes possible for the (partial) dual-unitary models. For definiteness, we choose the spatial direction marked by the index $n$ to be dual with the time direction, i.e. the evolution operator $u[f_v]$ with the symbol $f_v$ is unitary. The partial dual-unitarity assumption allows us to formulate the contraction rules for this spatial direction.
Indeed, by using  the unitarity of $u[f_v]$ the  part of the  sum (\ref{method1body}),
\begin{multline}\label{spatialcontraction}
\Gamma=L^{-1}\sum_{s=1}^L e^{- \i\big( f_v(s,\bar{s}_{n+1,m,t})- f_v(s,\ubar{s}_{n+1,m,t})\big)}
\\\times\delta(\bar{s}_{n,m+1,t},\ubar{s}_{n,m+1,t})\delta(\bar{s}_{n,m-1,t},\ubar{s}_{n,m-1,t})\\\times \delta(\bar{s}_{n,m,t-1},\ubar{s}_{n,m,t-1})\delta(\bar{s}_{n,m,t+1},\ubar{s}_{n,m,t+1})\\\times e^{- \i\big( f_h(s,\bar{s}_{n,m+1,t})- f_h(s,\ubar{s}_{n,m+1,t})\big)}\\\times  e^{- \i \big( f_h(\bar{s}_{n,m-1,t},s)- f_h(\ubar{s}_{n,m-1,t},s)\big)}\\
\times e^{- \i \big( g(\bar{s}_{n,m,t+1},s)- g(\ubar{s}_{n,m,t+1},s)\big)}\\\times e^{-\i\big( g(s,\bar{s}_{n,m,t-1})- g(s,\ubar{s}_{n,m,t-1})\big)}
\end{multline}
simplifies to 
\begin{equation}
    \Gamma= \delta(\bar{s}_{n+1,m,t},\underline{s}_{n+1,m,t}).
\end{equation} 
This allows us to formulate the contraction rule in the dual (spatial) direction, which is illustrated by the diagram in fig.~\ref{FigGrid1}a. It can be naturally extended to the opposite direction of the same spatial axis, see fig.~\ref{FigGrid1}b.

Note here that the contraction rule in the spatial direction becomes useful only for $N> T+2$. In this case, after subsequent applications of the contraction rules in the time direction, the border of the correlated spins is formed at some (at least one) given $n$. So that starting from this border we can subsequently apply the contraction rule in spatial direction.  

Application of the contraction rules both in spatial and temporal directions yields trivial correlations for the majority of choices of the operators $\bar{\Sigma}$ and $\underline{\Sigma}$  in eqs.~(\ref{barSigma}). All non-trivial cases result from a specific choice of $\q_\ell$ mutual positions. Namely, the operators $\q_\ell$ entering $\bar{\Sigma}$ and those from $\ubar{\Sigma}$ have to be positioned along the line given by the equation $\abs{n}=t, m=0$. A number of possible realisations are analysed in Appendix~\ref{AppB}. As it is demonstrated there, for the operators  $\bar{\Sigma}$, $\underline{\Sigma}$ with four-point supports, the non-trivial correlations arise only when $\mu=0$, $\nu=T+1$, see eq.~(\ref{barSigma}). An example of the spin structure resulting from the application of the contraction rules is depicted in fig.~\ref{Fig5}a. 

Finally, if duality holds for both spatial directions (fully dual unitary case), both $u[f_v]$ and $u[f_h]$ are unitary matrices. In this case, there is an additional set of contraction rules acting along the $m$-axis. Applying contraction rules along the third axis leads to the elimination of all spin variables (for $T>2$), regardless of the positions of $\bar{\Sigma}$ and $\ubar{\Sigma}$. This implies that $C(T)$ vanishes entirely when $T>2$.  

To summarize this section, we list the general properties of the correlation function  (\ref{tracesum1})  established so far for two-dimensional lattice models: (i) For a general (non-dual unitary) case the correlation function becomes trivial when the entries $\q_\ell$ ($\ell=5,6,7,8$) of the operator $\underline{\Sigma}$ are placed outside of the pyramid basis $\abs{n-3/2}+\abs{m-3/2}\le t+1$ (see fig.~\ref{Fig2}); (ii) In the partially dual unitary case, the correlation function becomes non-trivial when the operators $\bar{\Sigma}$ and $\underline{\Sigma}$ are aligned along the line $\abs{n}=t, m=0$ (see Appendix~\ref{AppB}). (iii) In the fully dual unitary case at $T>2$ the correlation function trivializes for any mutual positions of the operators $\q_\ell$.

\section{
The local correlation function
and the transfer operator.} \label{Sec6}

In this section, we show that similarly to the one-dimensional dual unitary case, the correlation function $C(T)$ in a two-dimensional partially dual unitary model can be expressed as an expectation value of the power of the low-dimension transfer operator (see eq.~\ref{CD1}).

\subsection{Operators with the two-point supports} 
Before addressing the correlation function for generic  operators $\bar{\Sigma}$, $\ubar{\Sigma}$ with four-point supports (see eq.~\ref{barSigma}), we illustrate our method for the case of the  operators with the two-point supports. To  this end we set in eqs.~(\ref{barSigma})  $\q_3=\q_4=\q_7=\q_8=\1$. An example of the structure of the uncorrelated spins obtained after applications of the contraction rules is illustrated in fig.~\ref{Fig4}a. In Appendix \ref{AppB} we list all non-trivial structures obtained after applications of the contraction rules. A part of the spin variables can be, furthermore, eliminated from the corresponding partition functions (see Appendix~\ref{AppB} for more details), so that only one non-trivial structure (up to the mirror reflection with respect to the plane $n=const$) survive. It is shown in fig.~\ref{Fig4}b. The reduced locus of spins, $\bm S'$, over which the summation still has to be performed (red, brown and green balls) is  
\begin{equation}
    \bm S' = \set{\bar{s}_{nmt} | (n,m,t) \in \mathcal{L}'}\cup\set{\ubar{s}_{nmt}| (n,m,t) \in \mathcal{L}''}
\end{equation}
with 
\begin{multline} 
\mathcal{L}'= 
\big\{(N,1,0),(1,M,0),(1,1,0),(1,2,0),(2,M,0),\\
(2,1,0),(2,2,0),(3,1,0),(1,1,1),\, 
(T,1,T),\\
(T+1,M,T),(T+1,1,T),(T+1,2,T),(T+2,M,T),\\
(T+2,1,T),(T+2,2,T),(T+3,1,T)\big\}\\
\bigcup_{t=1}^{T-1}\big\{(t+1,1,t),(t+2,1,t),(t+2,M,t),\\(t+2,2,t),(t+3,1,t)\big\},
\end{multline} 
\begin{multline} 
\mathcal{L}''=
\set{(1,1,0),(2,1,0),\,(T+1,1,T),(T+2,1,T)}\\\bigcup_{t=1}^{T-1}\set{(t+2,1,t)}.
\end{multline}

\begin{figure*}
\begin{tabular}{lccccc}
a)&\includegraphics[scale=0.25]{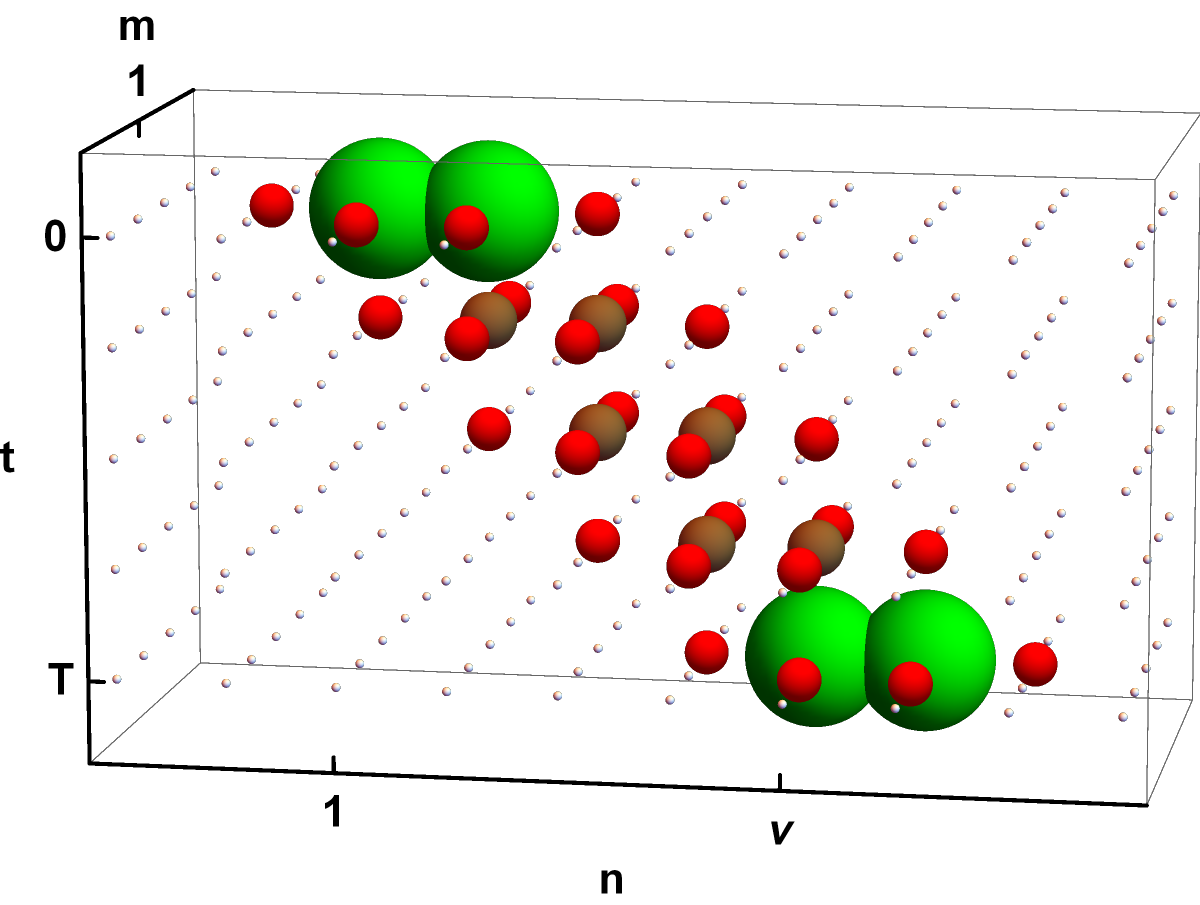}&
b)&\includegraphics[scale=0.25]{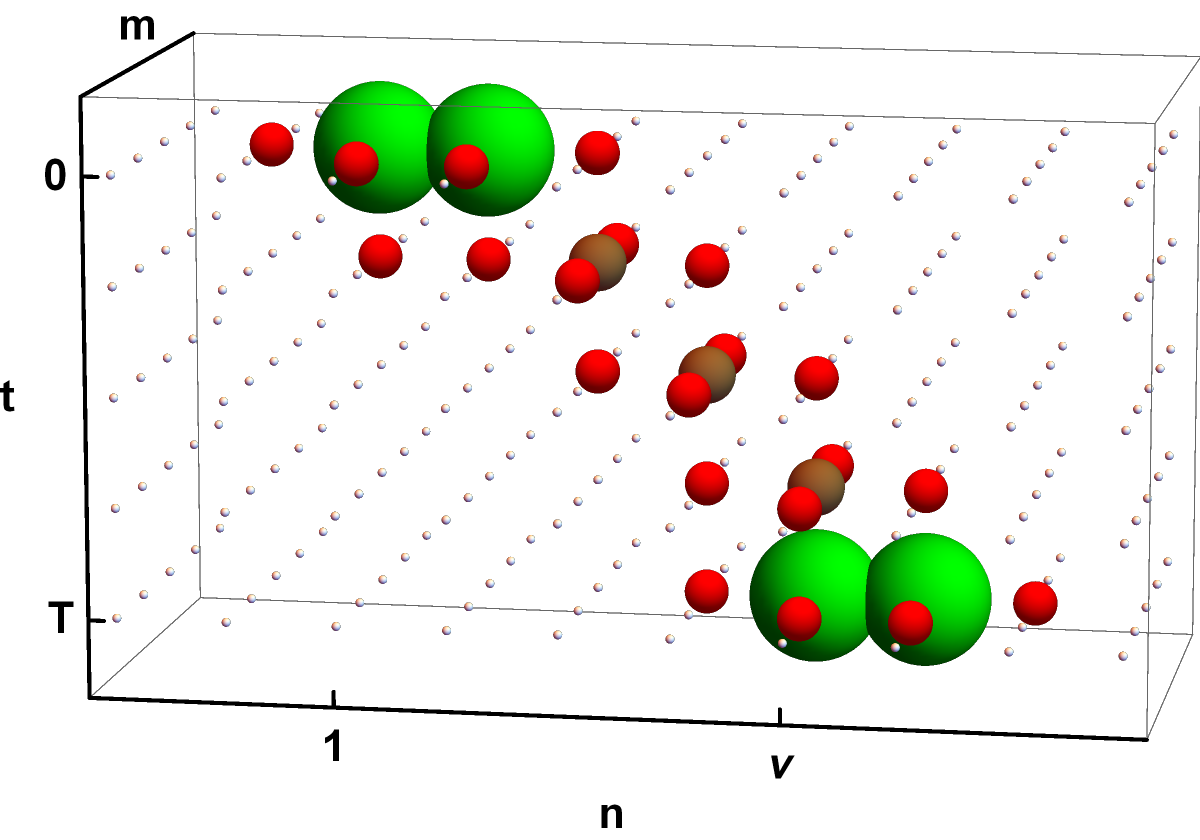}&
c)&\includegraphics[scale=0.16]{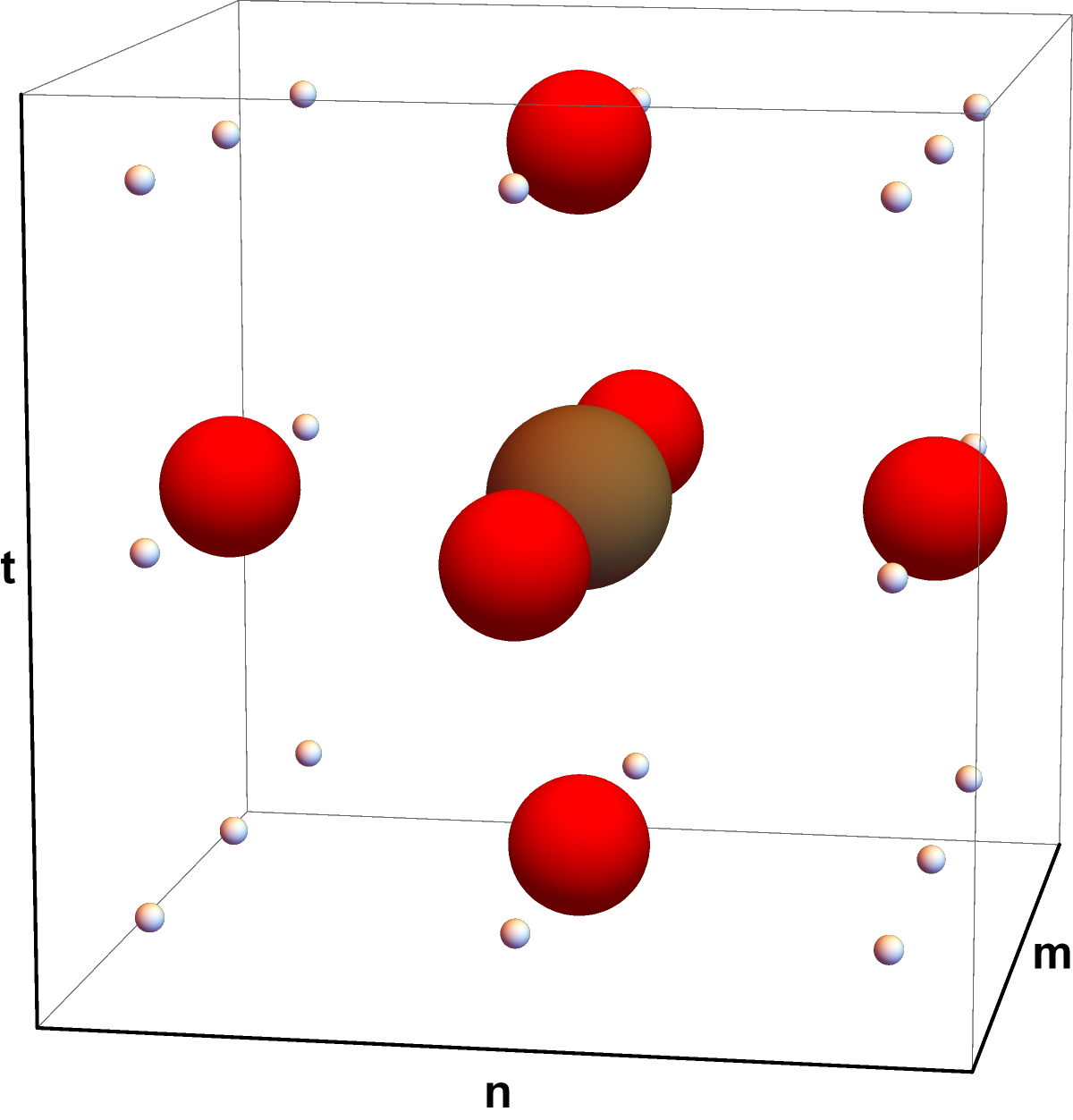}
\end{tabular}
\caption{\label{Fig4}\small a) The spin  structure obtained after application of the contraction rule to a partially dual unitary map for $T=4$, $M,N\gg T$ and $\mu=1,\nu=T+1$ in the case of operators with the two-point supports; b) The spin structure that emerges after eliminating a part of the spin variables (see Appendix~\ref{AppB} for details). The horizontal coordinates $n$ of the brown balls entering the intermediate linear structure of uncorrelated spins satisfy the equation $n=t+2$ and are made up of the repeating blocks (unit cells); c) The unit cell entering the spin-bridge structure corresponds to the transfer operator $\bm T$ (eq.~\ref{TD2}).} 
\end{figure*}

\begin{figure*}
\begin{tabular}{lccccc}
a)&\includegraphics[scale=0.25]{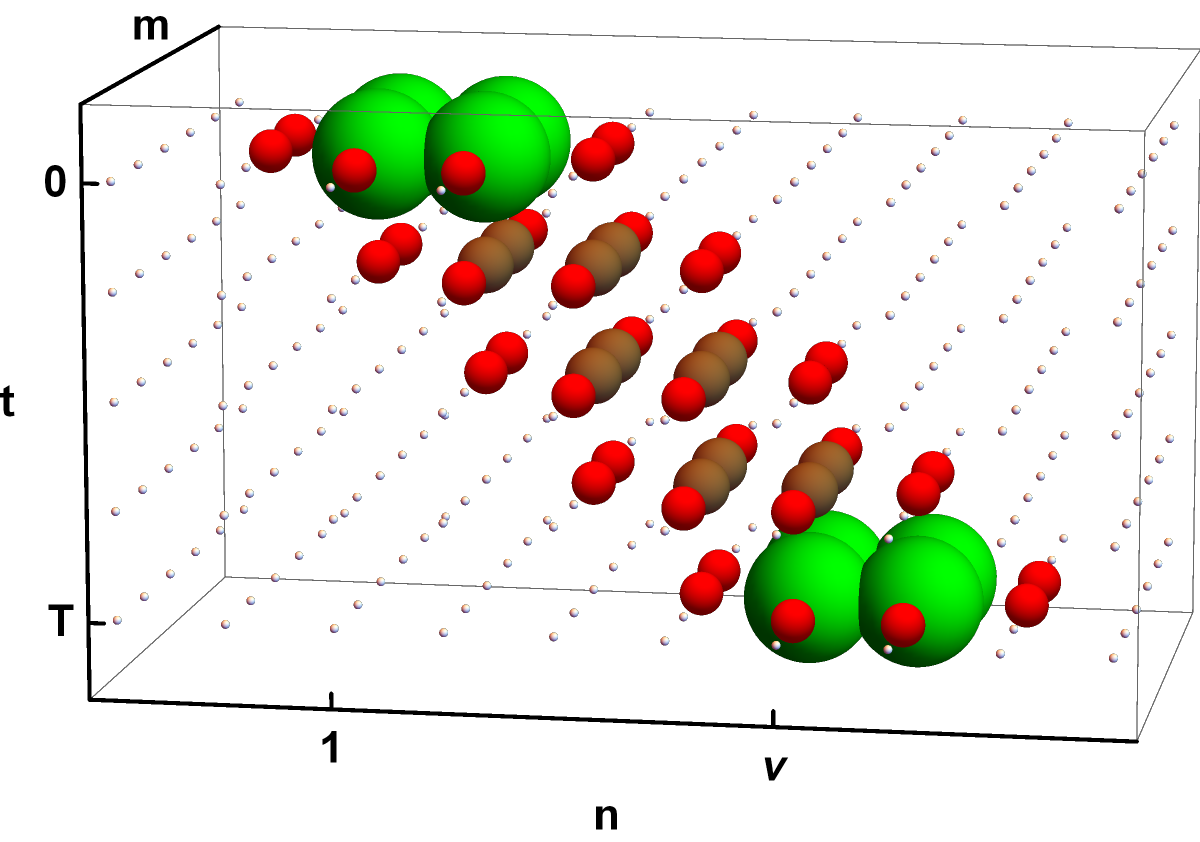}&
b)&\includegraphics[scale=0.25]{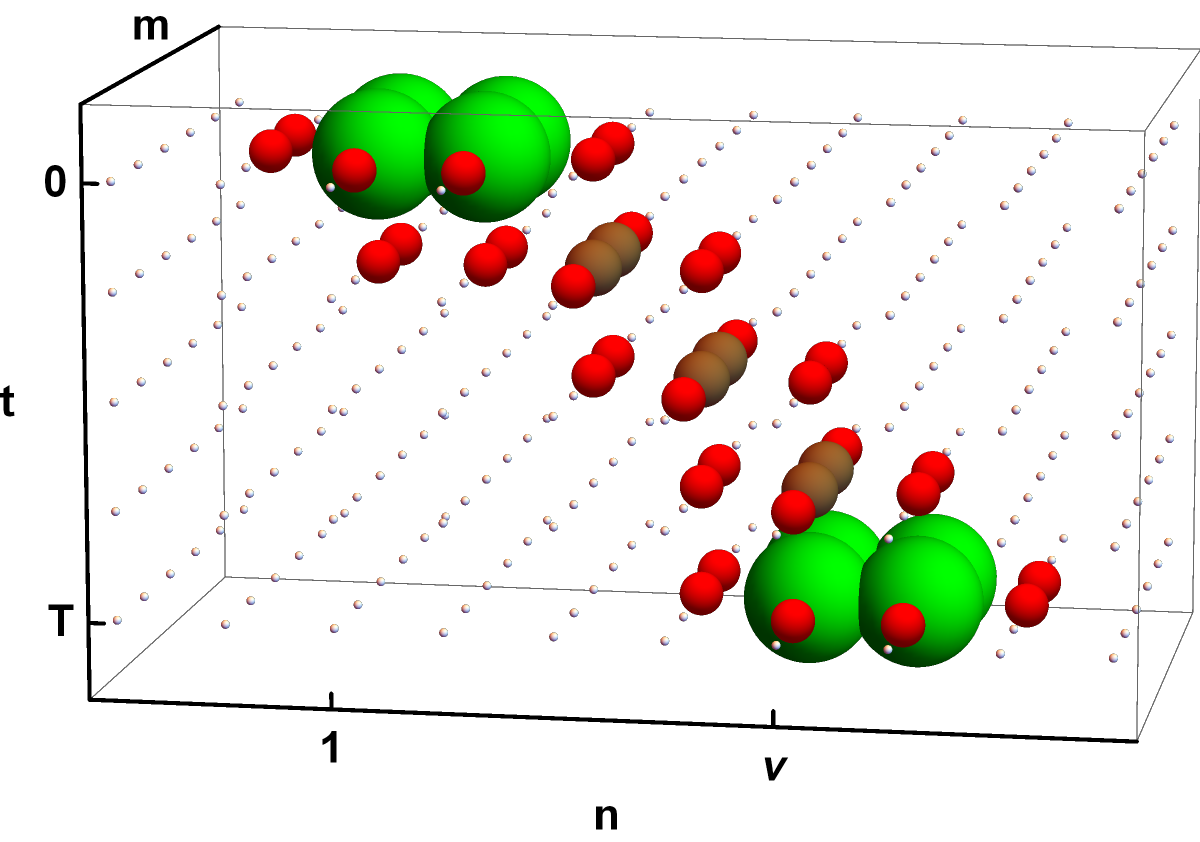}&
c)&\includegraphics[scale=0.16]{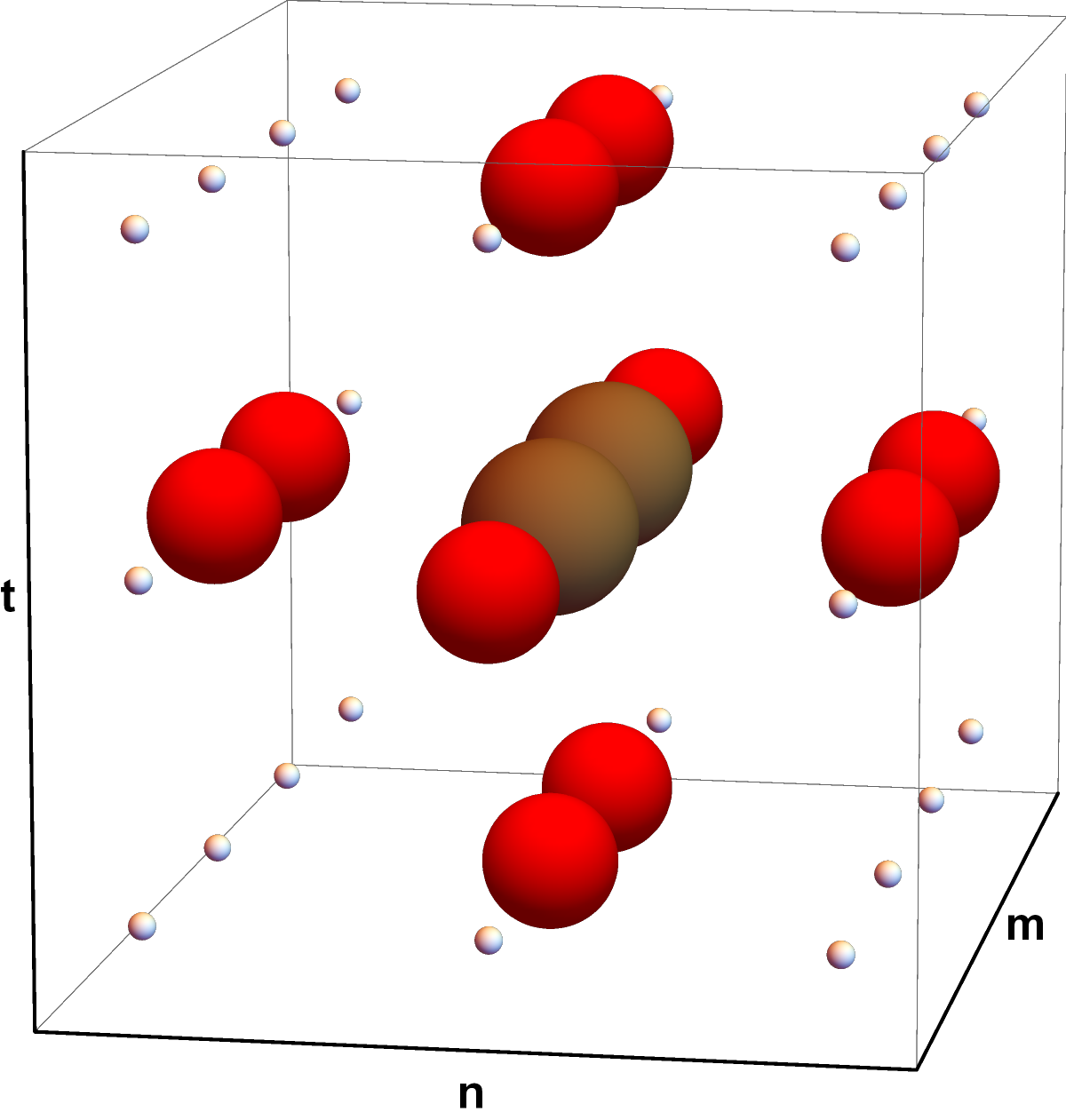}
\end{tabular}
\caption{\label{Fig5}\small a) The structure of spins obtained after application of the contraction rule to a partially dual unitary map for $T=4$, $M,N\gg T$ and $\mu=1, \nu=T+1$ corresponding to the non-trivial correlation function $C(T)$; b) The spin structure obtained from the one in fig.a being reduced by taking into account the boundary conditions (see Appendix~\ref{AppB} for details); c) The unit cell entering the linear structure of spins corresponds to the transfer operator $\bm T$ (eq.~\ref{TTD2}). } 
\end{figure*}

This chain of spin variables is composed of repeating blocks shown in fig.~\ref{Fig4}c, with their centres positioned along the straight line $n = t, m=0$. Since two neighbouring  unit cells are connected by a pair of spins, each block can be described by the  $L^2\times L^2$ transfer matrix $\bm T$ with  the entries $\braket{\chi,\eta|\bm T|\chi',\eta'}$, where  the indexes $\chi,\eta,\chi',\eta'=\overline{1,L}$ mark the  values of the spins to be convoluted with the spins of the neighbouring upper and lower unit cells. The matrix entries of $\bm T$ can be read off directly from the spin structure shown in fig.~\ref{Fig4}b, they are
\begin{multline} 
\braket{\chi,\eta|\bm T|\chi',\eta'}= \frac{1}{L^5}
\sum_{\bar{s},\ubar{s}=1}^L \sum_{r_1,r_2=1}^L e^{-\i f_h(r_1,\bar{s})+\i f_h(r_1,\ubar{s})}\\\times e^{-\i f_h(\bar{s},r_2)+ \i f_h(\ubar{s},r_2)}  e^{-\i f_v(\chi,\bar{s})+\i f_v(\chi,\ubar{s})-\i  g(\bar{s},\eta)+  \i g(\ubar{s},\eta)} \\
\times e^{-\i f_v(\bar{s},\eta')+ \i f_v(\ubar{s},\eta')-\i  g(\chi',\bar{s})+ \i g(\chi',\ubar{s})},
\end{multline}
The last expression can be written in a compact form
\begin{multline}\label{TD2}
\braket{\chi,\eta|\bm T|\chi',\eta'}
= \frac{1}{L^5}\sum_{r_1,r_2=1}^L
\\
\abs{\sum_{s=1}^L e^{\i f_v(\chi,s)+\i  g(s,\eta)+\i f_v(s,\eta') +\i  g(\chi',s)+\i f_h(r_1,s)+\i f_h(s,r_2)}}^2,
\end{multline}
The spin $\bar{s}$ and the conjugated spin $\ubar{s}$ in this expression  describe possible internal states of the central vertex in the unit cell (brown ball). The other two spin variables $r_1$ and $r_2$ describe the states of the neighbouring lattice points in the horizontal direction (red balls). It is worth noting that the matrix entries $\braket{\chi,\eta|\bm T|\chi',\eta'}$  coincide with those of the one-dimensional map (eq.~\ref{TD1}) when we set the horizontal interactions $f_h$ to zero. 
 
The overall expression for the correlation function is given by the expectation value, 
\begin{equation}\label{CT}
C(T)=\bra{\bar{\bm\Phi}_{\q_1,\q_2}}\bm T^{T-2}\ket{\bm \Phi_{ \q_5,\q_6}},
\end{equation}
where the entries of the transfer matrix $\bm T$ are given by eq.~(\ref{TD2}) and the vectors $\bra{\bar{\bm\Phi}_{\q_1,\q_2}}$,  $\ket{\bm \Phi_{\q_5,\q_6}}$ are defined below by eqs.~(\ref{Phi1}) -- (\ref{PhiG2}). 

The vector $\ket{\bar{\bm\Phi}_{\q_1,\q_2}}$ incorporates the function $\bar{\Phi}(\ubar{\bm s}_0,\bar{\bm s}_0)$, function $\bar{T}_I(\bar{\bm s}_0,\ubar{\bm s}_0)$ and a single phase drawn from the function $\bar{T}_K(\bar{\bm s}_1,\bar{\bm s}_1;\ubar{\bm s}_1,\ubar{\bm s}_0)$, so that
\begin{multline}\label{Phi1}
\braket{\bar{\bm\Phi}_{\q_1,\q_2}|\chi,\eta}=\\\frac{1}{L^5}\sum_{s_1,\bar{s}_2,\ubar{s}_2=1}^L\bar{ \Gamma}_{s_1}^{\bar{s}_2,\ubar{s}_2}(\chi,\eta)\bra{s_1}\q_1\ket{s_1}\bra{\bar{s}_2}\q_2\ket{\ubar{s}_2}
\end{multline}
with the factor $\bar{ \Gamma}_{s_1}^{\bar{s}_2,\ubar{s}_2}(\chi,\eta)$ given by
\begin{multline}\label{PhiG1}
\bar{ \Gamma}_{s_1}^{\bar{s}_2,\ubar{s}_2}(\chi,\eta)= e^{- \i  g(\chi,\bar{s}_2) +\i  g(\chi,\ubar{s}_2)}\\\times e^{- \i f_v(s_1,\bar{s}_2)+\i f_v(s_1,\ubar{s}_2)- \i f_v(\bar{s}_2,\eta)+\i f_v(\ubar{s}_2,\eta)}
\\\times\sum_{r_1,r_2=1}^L e^{-\i f_h(r_1,\bar{s}_2)+\i  f_h(r_1,\ubar{s}_2)-\i f_h(\bar{s}_2,r_2)+\i  f_h(\ubar{s}_2,r_2)}.
\end{multline}
The vector $\ket{\bm \Phi_{\q_5,\q_6}}$ incorporates two time slices, namely it incorporates the functions $\Phi(\bar{\bm s}_T,\ubar{\bm s}_T)$, $\bar{T}_I(\bar{\bm s}_T,\ubar{\bm s}_T)$, $\bar{T}_K(\bar{\bm s}_{T-1},\bar{\bm s}_T;\ubar{\bm s}_T,\ubar{\bm s}_{T-1})$ and a single phase factor drawn from the function $\bar{T}_K(\bar{\bm s}_{T-2},\bar{\bm s}_{T-1};\ubar{\bm s}_{T-1},\ubar{\bm s}_{T-2})$. Note, that the contraction rules were formulated in a symmetric manner, namely, we have introduced the additional matrices $U_I$ into the functions $\Phi(\bar{\bm s}_T,\ubar{\bm s}_T)$, so that the product $\Phi(\bar{\bm s}_T,\ubar{\bm s}_T)\bar{T}_I(\bar{\bm s}_T,\ubar{\bm s}_T)$ naturally simplifies to the scalar product $\bra{\bar{\bm s}_T}\ubar{\Sigma}\ket{\ubar{\bm s}_T}$. The entries $\braket{\chi',\eta'|\bar{\bm\Phi}_{\q_5,\q_6}}$ are defined by the expression
\begin{multline}\label{Phi2}
\braket{\chi',\eta'|\bm\Phi_{\q_5,\q_6}}=\frac{1}{L^5}\sum_{\bar{s}_1,\ubar{s}_1,s_2=1}^L \Gamma_{\bar{s}_1,\ubar{s}_1}^{s_2}(\chi',\eta')\\ \times
\bra{\bar{s}_1}u^\dag[g]\q_5u[g]\ket{\ubar{s}_1}\bra{s_2}u^\dag[g]\q_6u[g]\ket{s_2}
\end{multline}
with 
\begin{multline}\label{PhiG2}
\Gamma_{\bar{s}_1,\ubar{s}_1}^{s_2}(\chi',\eta')=e^{- \i  g(\bar{s}_1,\eta') +\i  g(\ubar{s}_1,\eta')}\\\times e^{- \i f_v(\chi',\bar{s}_1)+\i f_v(\chi',\ubar{s}_1)- \i f_v(\bar{s}_1,s_2)+\i f_v(\ubar{s}_1,s_2)} \\\times\sum_{r_1,r_2=1}^L e^{-\i f_h(r_1,\bar{s}_1)+\i  f_h(r_1,\ubar{s}_1)-\i f_h(\bar{s}_1,r_2)+\i  f_h(\ubar{s}_1,r_2)}.
\end{multline}

The expression  (\ref{CT}) was obtained for the structure shown in fig.~\ref{Fig4}a, when the spin-bridge structure is parallel to the line $n=t$. The  correlation function,
\begin{equation}\label{CT1}
C(T)=\bra{\bar{\bm\Phi}^r_{\q_1,\q_2}} {\bar{\bm T}}^{T-2}\ket{\bm \Phi^r_{\q_5,\q_6}},
\end{equation}
for the mirror-symmetric structure of spins (along the line $n=-t$) can be obtained by using the symmetry arguments. 
The ``reflected'' transfer matrix $\bar{\bm T}$ has the entries $\braket{\eta',\chi'|\bar{\bm T}|\eta,\chi}=\braket{\eta',\eta|\bm T|\chi',\chi}$ and the boundary vectors for the reflected picture are
\begin{multline}\label{PhiM1}
\braket{\eta,\chi|\bar{\bm\Phi}^{r}_{\q_1,\q_2}}=\frac{1}{L^5}\sum_{\bar{s}_1,\ubar{s}_1,s_2=1}^L\bar{ \Gamma}^{s_2}_{\bar{s}_1,\ubar{s}_1}(\eta,\chi)\\\times \bra{\bar{s}_1}\q_1\ket{\ubar{s}_1}\bra{s_2}\q_2\ket{s_2};\end{multline}
\begin{multline}
\braket{\bm\Phi_{\q_5,\q_6}^{r}|\eta',\chi'}=\frac{1}{L^5}\sum_{s_1,\bar{s}_2,\ubar{s}_2=1}^L \Gamma^{\bar{s}_2,\ubar{s}_2}_{s_1}(\eta',\chi')\\\times \bra{s_1}u^\dag[g]\q_5u[g]\ket{s_1}\bra{\bar{s}_2}u^\dag[g]\q_6u[g]\ket{\ubar{s}_2}\label{PhiM2}
\end{multline}
with 
\begin{multline}
\bar{ \Gamma}^{s_2}_{\bar{s}_1,\ubar{s}_1}(\eta,\chi)= e^{- \i  g(\chi,\bar{s}_1) +\i  g(\chi,\ubar{s}_1)}
\\\times e^{- \i f_v(\bar{s}_1,s_2)+\i f_v(\ubar{s}_1,s_2)- \i f_v(\eta,\bar{s}_1)+\i f_v(\eta,\ubar{s}_1)}
\\\times \sum_{r_1,r_2=1}^L e^{-\i f_h(r_1,\bar{s}_1)+\i  f_h(r_1,\ubar{s}_1)-\i f_h(\bar{s}_1,r_2)+\i  f_h(\ubar{s}_1,r_2)};\label{PhiGM1}
\end{multline}
\begin{multline}\label{PhiGM2}
\Gamma^{\bar{s}_2,\ubar{s}_2}_{s_1}(\eta',\chi')=e^{- \i  g(\bar{s}_2,\eta') +\i  g(\ubar{s}_2,\eta')}\\\times e^{- \i f_v(\bar{s}_2,\chi')+\i f_v(\ubar{s}_2,\chi')- \i f_v(s_1,\bar{s}_2)+\i f_v(s_1,\ubar{s}_2)}\\\times\sum_{r_1,r_2=1}^L e^{-\i f_h(r_1,\bar{s}_2)+\i  f_h(r_1,\ubar{s}_2)-\i f_h(\bar{s}_2,r_2)+\i  f_h(\ubar{s}_2,r_2)}.
\end{multline}

\subsection{Operators with the four-point supports} 
For generic operators  $\bar{\Sigma}$, $\ubar{\Sigma}$ with four-point supports (see eq.~\ref{barSigma}),
the  spin structure emerging after the application of the contraction rules gets the structure shown in fig.~\ref{Fig5}b. Each unit cell (fig.~\ref{Fig5}c) composing the bridge between the boundaries (green balls) is  described now by the  $L^4\times L^4$ transfer matrix $\bm T$ with the entries
\begin{multline}\label{TTD2}
\braket{\chi,\eta,\chi_1,\eta_1|\bm T|\chi',\eta',\chi'_1,\eta'_1}= \frac{1}{L^7}\sum_{r_1,r_2=1}^L 
\\
\times \Bigg| \sum_{s_1,s_2=1}^L  e^{\i f_v(\chi,s_1)+  \i g(s_1,\eta) + \i f_v(s_1,\eta')+ \i g(\chi',s_1)}
\\ \times e^{\i f_h(r_1,s_1)+ \i f_h(s_1,s_2)+\i f_h(s_2,r_2)} 
\\\times
e^{\i f_v(\chi_1,s_2) +  \i g(s_2,\eta_1)+ \i f_v(s_2,\eta'_1)+ \i g(\chi'_1,s_2)}\Bigg|^2.
\end{multline}
Note, that  after summation over the indexes $\chi_1,\eta_1$, the matrix element
$\braket{\chi,\eta,\chi_1,\eta_1|\bm T|\chi',\eta',\chi'_1,\eta'_1}$ reduces to the one in eq.~(\ref{TD2}).
As in the  case of the operators with two-point supports, the correlation function is given by the expectation value, 
\begin{equation}\label{CT2}
C(T)=\bra{\bar{\bm\Phi}_{\bar{\Sigma}}}\bm T^{T-2}\ket{\bm \Phi_{\ubar{\Sigma} }},
\end{equation} with the vectors $\bra{\bar{\bm\Phi}_{\bar{\Sigma}}}$,  $\ket{\bm \Phi_{\ubar{\Sigma}}}$  determined by the operators $\q_1,\q_2,\q_3,\q_4$ and $\q_5,\q_6,\q_7,\q_8$, respectively. The explicit expressions for these vectors are quite cumbersome, and we do not provide them here.

\subsection{Spectral properties of the transfer operator}

It follows immediately from the unitarity of matrices $u[g]$ and $u[f_v]$ that  the transfer matrix (\ref{TTD2}), as well as  its reduced form (\ref{TTD2})), is  doubly stochastic, i.e. it satisfies the property 
\begin{eqnarray}\label{doublystocahsticT}
\sum_{\chi_1,\eta_1,\chi,\eta=1}^L\braket{\chi,\eta,\chi_1,\eta_1|\bm T|\chi',\eta',\chi'_1,\eta'_1}
=1&&\nonumber\\ 
\sum_{\chi'_1,\eta'_1,\chi',\eta'=1}^L
\braket{\chi,\eta,\chi_1,\eta_1|\bm T|\chi',\eta',\chi'_1,\eta'_1}
=1&&.
\end{eqnarray}
 The  identity (\ref{doublystocahsticT}), in particular, means that the spectrum of $\bm T$ is contained within the unit disc on the complex plane with the largest eigenvalue $\lambda_0=1$. The eigenvector corresponding to the maximal eigenvalue has the    constant entries,
\begin{equation*} \braket{\mathcal{E}|\chi,\eta,\chi_1,\eta_1}=1, \quad\chi,\eta,\chi_1,\eta_1\in \overline{1,L}.
\end{equation*}
It is straightforward to see that for traceless operators $\bar{\Sigma}, \ubar{\Sigma}$ both 
$\bra{\bar{\bm\Phi}_{\bar{\Sigma}}}$ and $\ket{\bm \Phi_{\ubar{\Sigma}}}$  are orthogonal to  $\mathcal{E}$,
 \begin{equation*}
\braket{\bar{\bm\Phi}_{\bar{\Sigma}} |\mathcal{E}} = \braket{\mathcal{E}|\bm \Phi_{\ubar{\Sigma}}}=0.
 \end{equation*}
As a result, for the traceless operators, the leading  contribution into the correlation function (\ref{CT2}), (\ref{CT}) is determined by  the second largest eigenvalue, $\lambda_1$ ($\abs{\lambda_1}<\abs{\lambda_0}$), of the operator $\bm T$. 
Thus, generically,  the non-trivial part of the correlation function decays exponentially with time and the characteristic decay rate is $|\ln \abs{\lambda_1}|^{-1}$. It is instructive, therefore, to study the dependence of $\lambda_1$ on the internal parameters of the model. In the next section, we provide an analysis of the transfer matrix spectrum  for a particular choice of the  map.

\section{Applications}\label{Sec7}

In this section, we illustrate our results using two particular realizations of the general model: coupled cat maps and  kicked Ising spin-lattice. Specifically, we provide  a detailed spectral analysis of the transfer matrix  (\ref{TD2}), for the case of operators with two-point supports.

\subsection{Coupled  cat maps} One of the best studied and understood examples of systems with chaotic dynamics is provided by Arnold's cat map, which is the hyperbolic automorphism of the two-dimensional unit torus~\cite{ArnoldBook}. The cat map acts in the 2D  phase space: $\set{x_t,p_t}\to\set{x_{t+1},p_{t+1}}$, with  $x_t,p_t$ being the coordinate and momentum at the discreet moment of time  $t$. The generation function of a single perturbed cat map is the function
\begin{equation}\label{SAction}
\mathcal{S}(x_t,x_{t+1})=\frac{1}{2}\left(ax_t^2 +2c x_t x_{t+1} +  b x_{t+1}^2\right) +\mathcal{V}(x_t),
\end{equation}
where $a,b,c$ are integers and 
$\mathcal{V}(x)$ is an arbitrary smooth real-valued function satisfying the periodic conditions, $\mathcal{V}(x+1)=\mathcal{V}(x)$. The equations of motion are defined through the derivatives of the action,  
$x_t=\partial \mathcal{S}/\partial x_{t+1}$, $p_t=-\partial \mathcal{S}/\partial x_t$. Their explicit form is 
\begin{equation}\label{CatMapTransf}
 \begin{cases}x_{t+1}= a x_t+p_t +\mathcal{V}'(x_t),& \mod 1;\\
p_{t+1}=  (a b-1) x_t+ b p_t ,& \mod 1,
\end{cases}
\end{equation}
where we set $c=-1$ for the sake of simplicity of exposition.  The regime of fully  chaotic dynamics is achieved when $a+b>2$. The above dynamical equations can be cast into a more compact, Newton form:
\begin{equation}\label{CatMapNewton}
  x_{t-1}+  x_{t+1}=(a+b)x_t +\mathcal{V}'(x_t) \mod 1.
\end{equation}

An extension of the cat map  to the  many-body setting, \textit{coupled cat map lattice},  was introduced  in~\cite{GutOsi15}   and subsequently studied in a number of works~\cite{GHJSC16, LiangCvit20022, Fouxon_Gutkin_2022} both on classical and quantum levels.
In  this  model, $\mathcal{N}$ cat maps, placed on the sites of the $D$-dimensional lattice $\Lat$, are coupled with the help of nearest-neighbour linear interactions. The resulting dynamical equations for $D=2$ take the form
\begin{multline}
d_h\left(x_{n+1,m, t}+x_{n-1,m, t}\right)+d_v\left(x_{n,m+1, t}+x_{n,m-1, t}\right)\!=\\
=x_{n,m, t+1}+x_{n,m, t-1}-(a+b) x_{n,m, t}
-\mathcal{V}'(x_{n,m, t})\mod 1, \label{coupledcats}
\end{multline}
where $x_{n,m, t}$ stands for the cat's coordinate  at the $(n,m)$-site of the lattice.
The constants  $d_h,d_v$ in   eq.~\ref{coupledcats} determine the strength of the coupling in the horizontal and vertical directions, respectively.  The model is partially dual-unitary if one of the coupling constants equals  $-1$ and  fully dual-unitary when $d_h=d_v=-1$. Indeed, as can be readily observed, the  eq.~\ref{coupledcats} remains  invariant under the exchange of $t$ and $n$ if $d_h=-1$,  or under the exchange of  $t$ and $m$ if $d_v=-1$.  Since we are primarily interested in the partially dual-unitary case, we fix $d_v=-1$ from now on and leave $d_h$ as a free parameter.

The quantisation of a single cat map  can be carried out according to a general procedure for  quantisation   of linear automorphism, see \cite{HannayBerry1980,  Rivas_2000}. The corresponding unitary  time evolution is given by $L\times L$ matrix $u[g]$ of the form (\ref{usmall}), where the function $g$ is determined by the classical action (\ref{SAction}), 
\begin{equation}\label{g1}
    g(s,s')=\frac{2\pi}{L}\mathcal{S}(s,s').
\end{equation}
Note that $u[g]$ is a Hadamard matrix, with  the factor $2\pi/L$ playing the role of the effective Planck's constant. 
An extension of this quantization procedure to coupled cat map lattice was presented in \cite{ Fouxon_Gutkin_2022}. In accordance with the structure of the   classical map, the   corresponding   quantum  evolution  can be split into  the product, $U=U_K[g] U_I[f_h,f_v]$, where $U_K$ is given by the tensor product of $\N$ operators $u[g]$ and  $U_I[f_h,f_v]$ is an interaction part provided by the diagonal matrix  (\ref{Uinter}), with
\begin{equation}
f_v(s,s')= \frac{2\pi}{L} ss', \qquad
f_h(s,s')= \frac{2\pi}{L}d_h s s'.\label{fh1}
\end{equation}

\begin{figure*} 
\begin{tabular}{lll}
\includegraphics[scale=0.35]{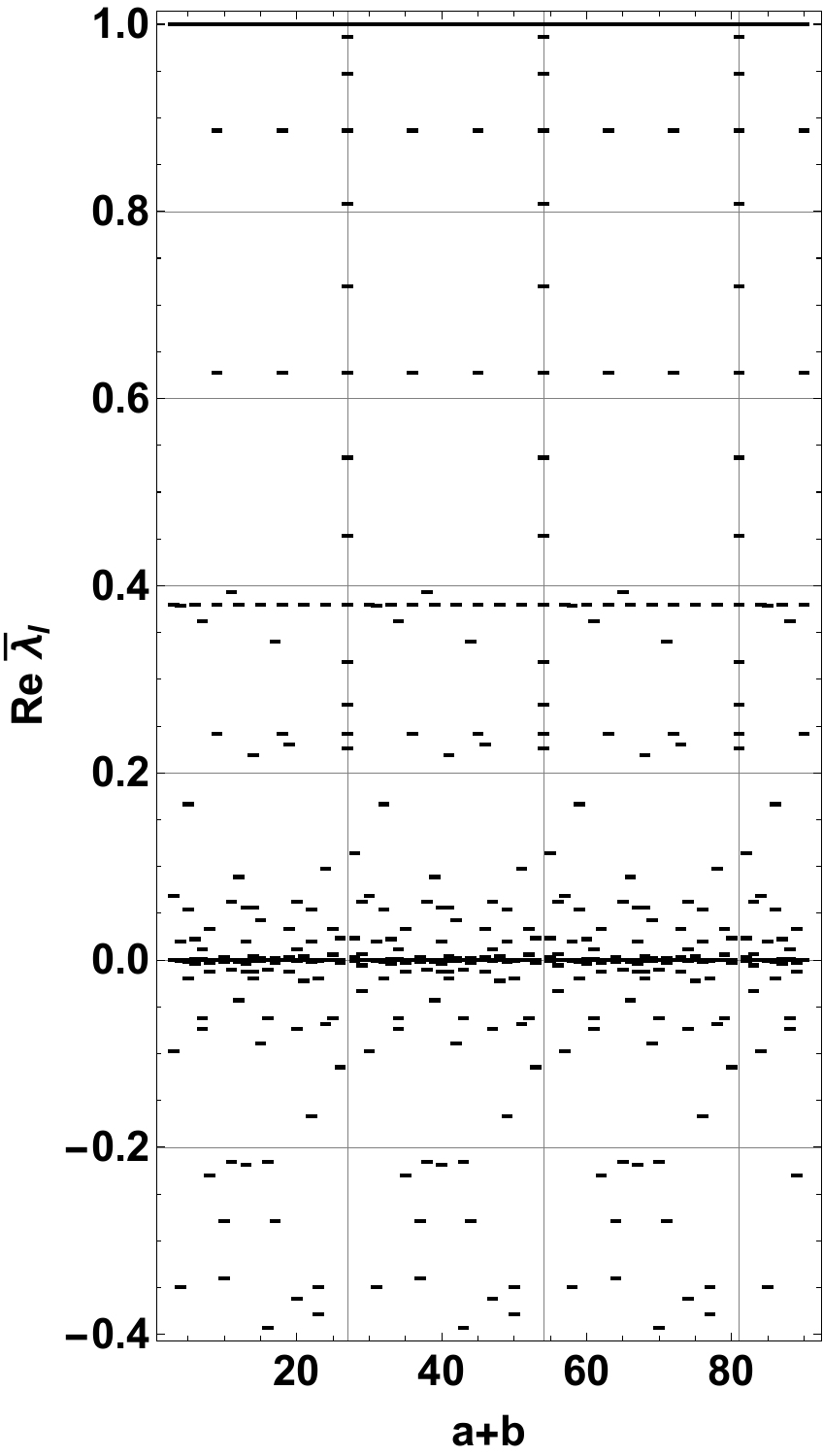}&\includegraphics[scale=0.35]{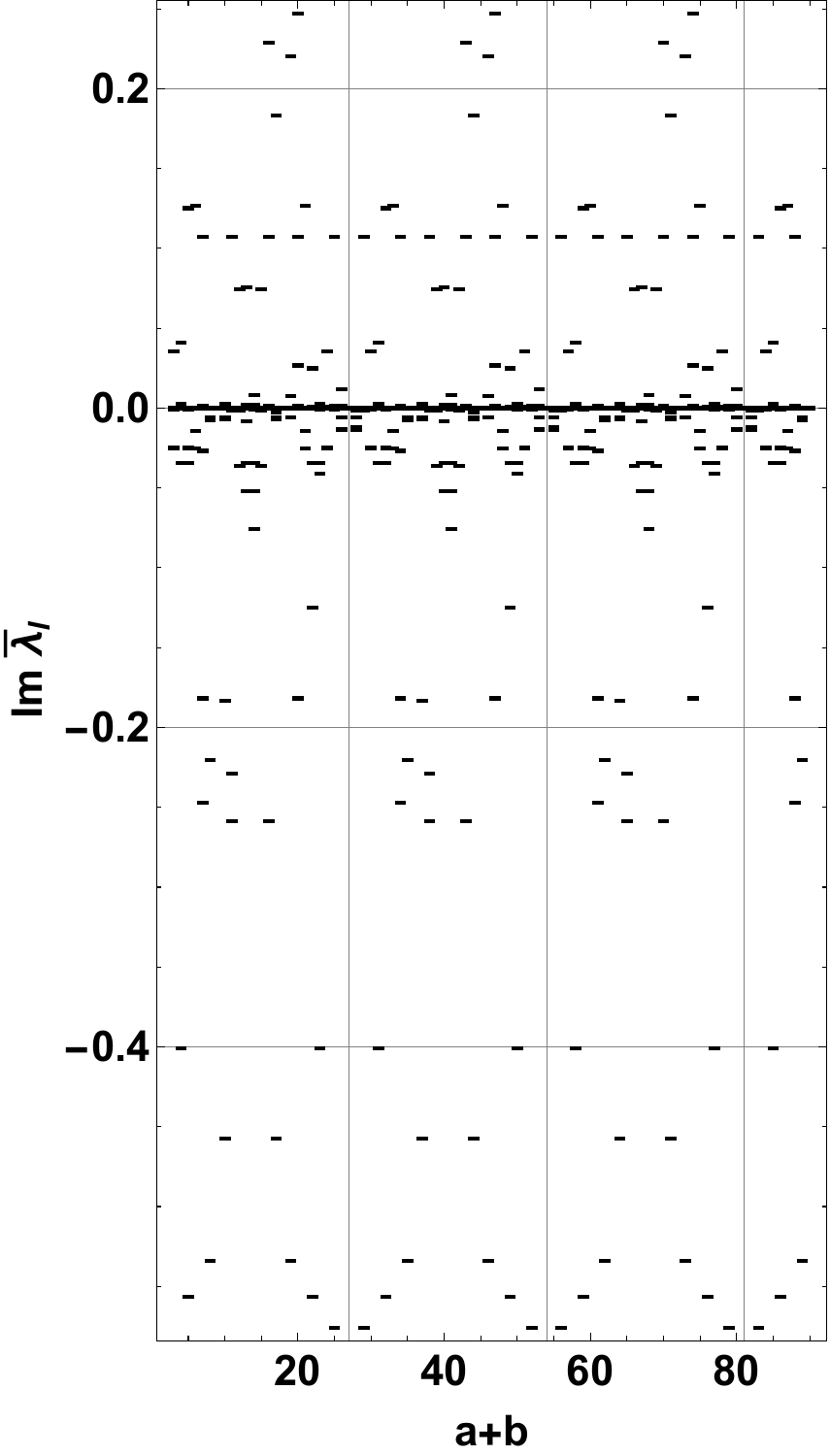}&\includegraphics[scale=0.35]{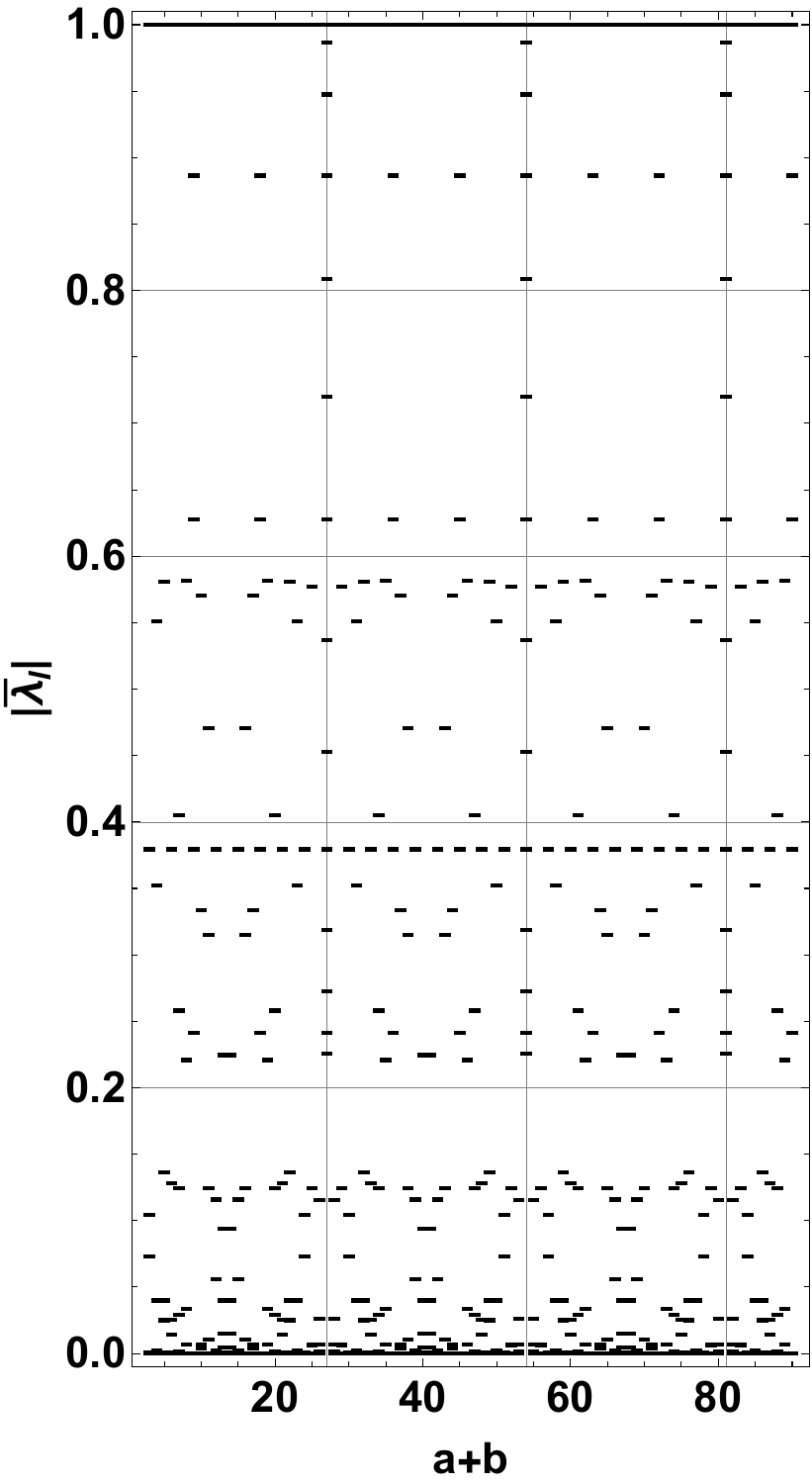}
\end{tabular}
\caption{\label{Fig7}\small Real part, imaginary part and the absolute value of the eigenvalues $\bar{\lambda}_{\ell}$ (eq.~\ref{lambdabar}) v.s. $a+b$ at $L=27$ and $\mathcal{V}(s)=\cos 2\pi s$. The eigenvalues change periodically with respect to $a+b$ with the period $L$.} 
\end{figure*}

\begin{figure*} 
\begin{tabular}{lll}
\includegraphics[scale=0.35]{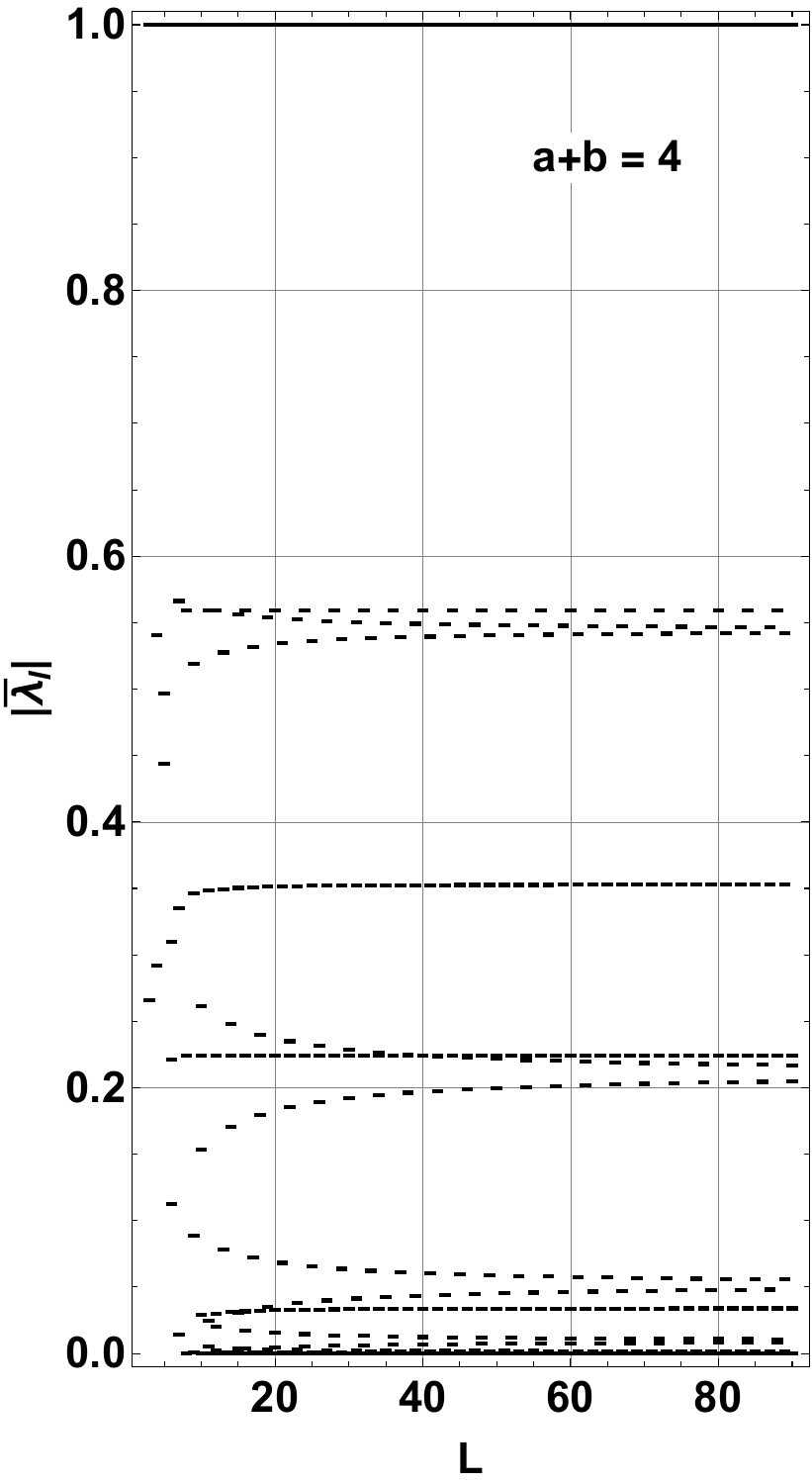}&
\includegraphics[scale=0.35]{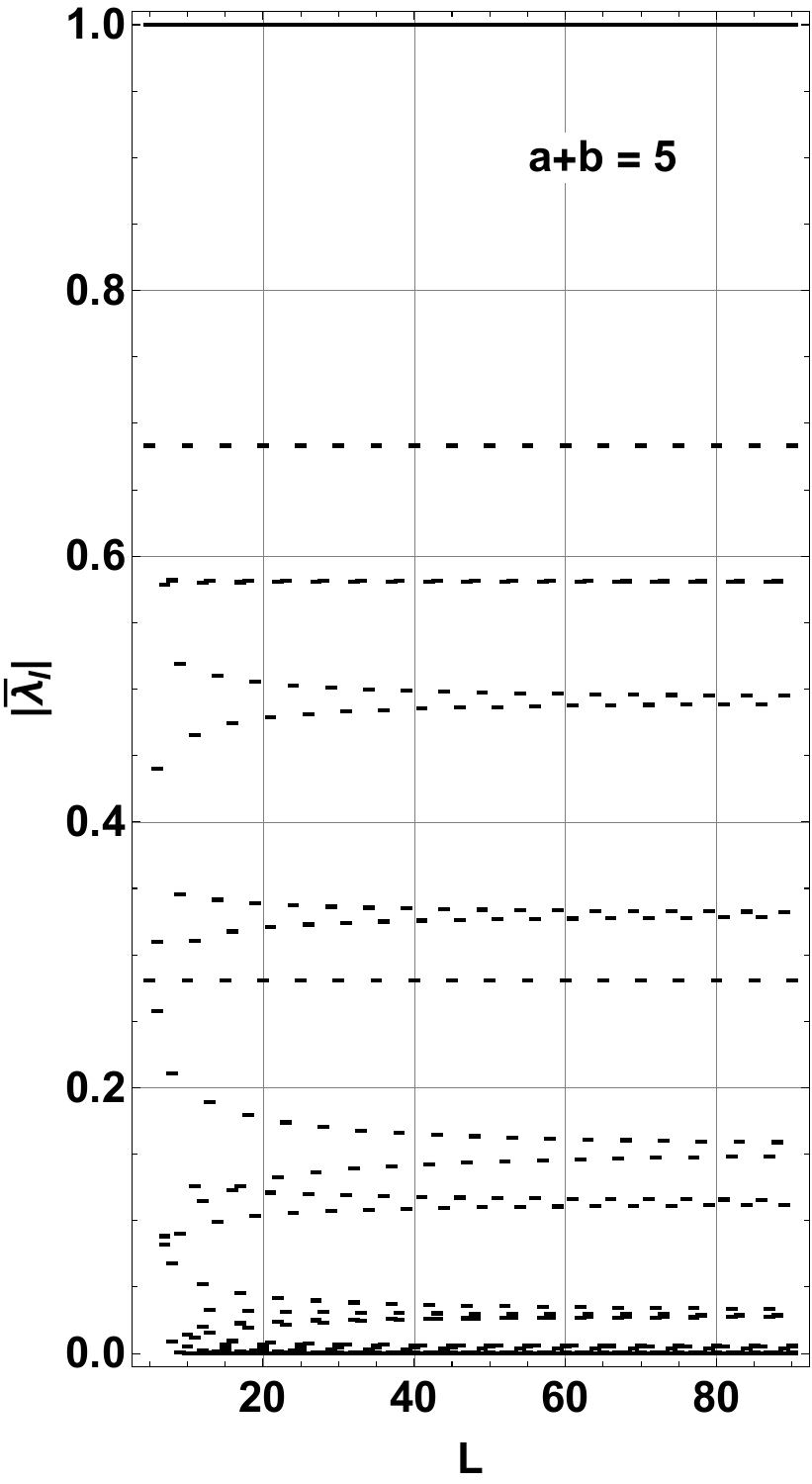}&
\includegraphics[scale=0.35]{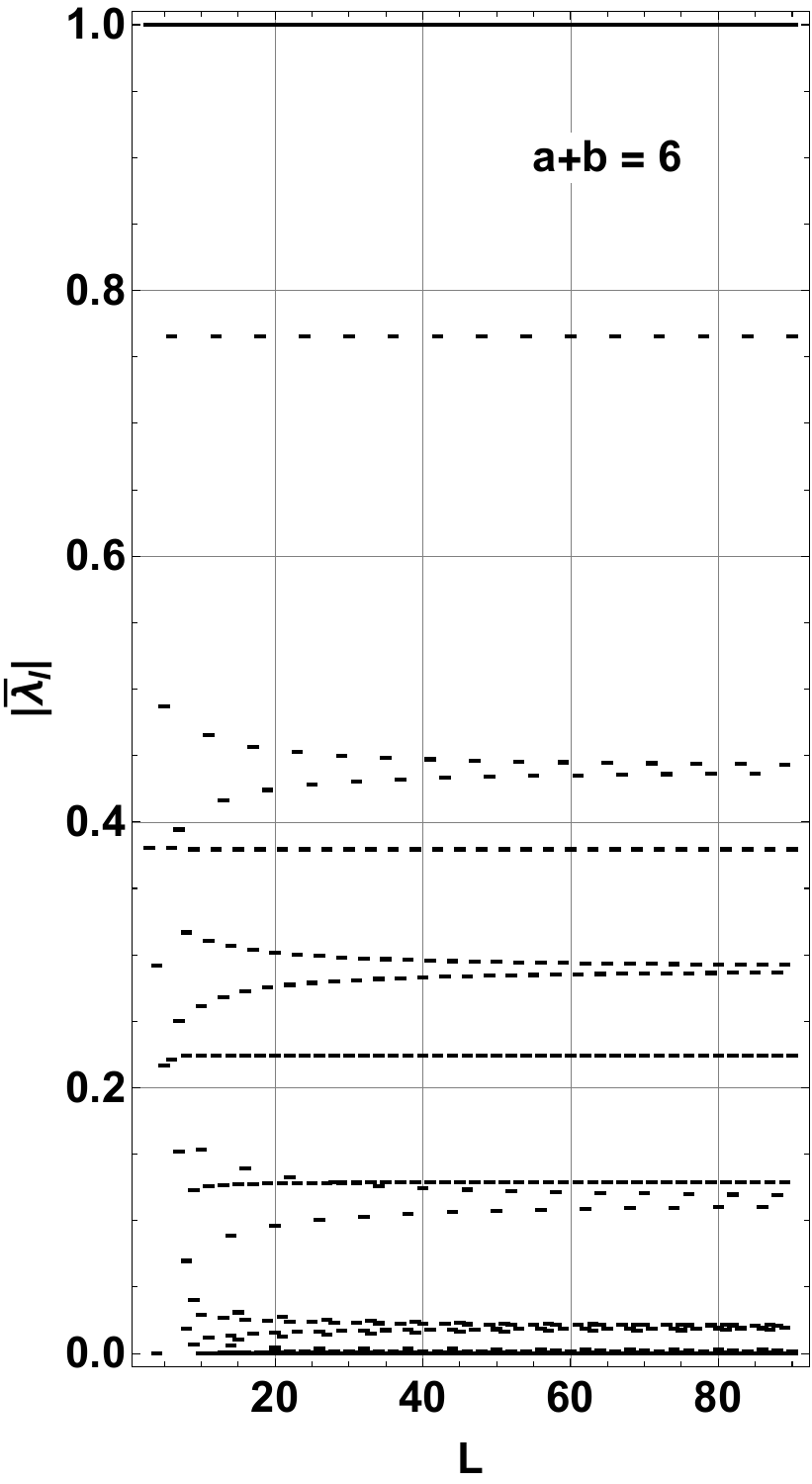}
\end{tabular}
\caption{\label{Fig8}\small The spectrum $\abs{\bar{\lambda}_{\ell}}$ (eq.~\ref{lambdabar}) of the one-dimensional cat-map model transition operator plotted with respect to the particle Hilbert space dimension $L$ is plotted for three different values of the parameter $a+b$. The perturbation is $\mathcal{V}(s)=\cos 2\pi s$.} 
\end{figure*}

Since the resulting time evolution $U$ is partially  dual-unitary and possesses the required  form (\ref{Unitgeneric}), we can straightforwardly apply the results from Sec.~\ref{Sec6}. For the above set of functions $g,f_v,f_h$   the transfer matrix entries (eq.~\ref{TD2}) become
\begin{multline}\label{TDCM}
\braket{\chi,\eta|\bm T|\chi',\eta'}
= \frac{1}{L^5}
\sum_{r_1,r_2=1}^L \\ 
\abs{\sum_{s=1}^L e^{\i \frac{2\pi}{L} s( \chi- \eta +\eta' - \chi') +\i \frac{\pi}{L}(a+b)s^2 +\i \mathcal{V}(s/L) +\i \frac{2\pi}{L}d_h s(r_1+r_2)}}^2.
\end{multline} 
A brief analysis of the matrix elements (eq.~\ref{TDCM}) immediately shows that, since the matrix indexes enter in the combination $(\chi- \eta) - (\chi'-\eta')$, among all $L^2$ matrix rows only $L$ rows are linearly independent. Therefore, for each choice of $L$ the transfer matrix has only $L$ non-zero eigenvalues. The non-trivial kernel of the transfer matrix is the Toeplitz matrix, i.e. the matrix entries depend on the difference of their indexes. It has the entries $K_{j-k}$ ($j,k=\overline{0,L-1}$), with
\begin{multline}
K_{j}= \frac{1}{L^5}\sum_{r_1,r_2=1}^L
\\ \abs{\sum_{s=1}^L e^{\i \frac{2\pi}{L}sj +\i \frac{\pi}{L}  (a+b)s^2 +\i \mathcal{V}(s)+\i \frac{2\pi}{L}d_h s(r_1 + r_2)}}^2.
\end{multline}
The Toeplitz matrices are known to be diagonalizable by the Fourier matrix with the entries $F_{k,\ell}=L^{-1/2}\exp[ 2\pi\i\, k \ell /L]$,
namely
\begin{eqnarray}
\sum_{j,k=0}^{L-1} F^*_{\ell,j}K_{j-k} F_{k,\ell'} 
&=&\delta(\ell-\ell')\lambda_\ell;\\
\lambda_{\ell}&=&\sum_{j=0}^{L-1} e^{-\i\frac{2\pi}{L} \ell j} K_j,\label{EigenvalueTrOp}
\end{eqnarray}
where $\ell$ runs from $0$ to $L-1$.
After performing the summation in eq.~(\ref{EigenvalueTrOp}), the non-trivial eigenvalues $\lambda_\ell$ of $T$ can be written in a compact form
\begin{eqnarray}
\label{lambdaohnebar}
&&\lambda_{\ell}=\bar{\lambda}_{\ell}e^{\i \frac{\pi\ell}{L}\left[(a+b)\ell+2 (L+1) d_h\right]}R_{L,\ell} (d_h),\\ \label{Rfunction} &&R_{L,\ell}(d_h)=\frac{\sin^2\pi d_h\ell}{L^2\sin^2\frac{\pi d_h\ell}{L}},
\end{eqnarray}
where the first factor,
\begin{equation}\label{lambdabar}
\bar{\lambda}_{\ell}=\frac{1}{L}\sum_{s=1}^L e^{\i \frac{2\pi}{L}  (a+b)s\ell +\i \mathcal{V}(s/L)-\i \mathcal{V}(s/L+\ell/L)},
\end{equation}
   represents  eigenvalues of the transfer matrix for the dual-unitary couple cat map chain   ($D=1$).  The real and the imaginary parts, as well as the absolute value of $\bar{\lambda}_\ell$,  are plotted in fig.~\ref{Fig7} for $L=27$ and the perturbation $\mathcal{V}(s)=\cos 2\pi s$.
The plot showing dependence of $\bar{\lambda}_\ell$ on $L$ is given in fig.~\ref{Fig8}.

\begin{figure*}
\begin{tabular}{lccc}
a)&\includegraphics[scale=0.40]{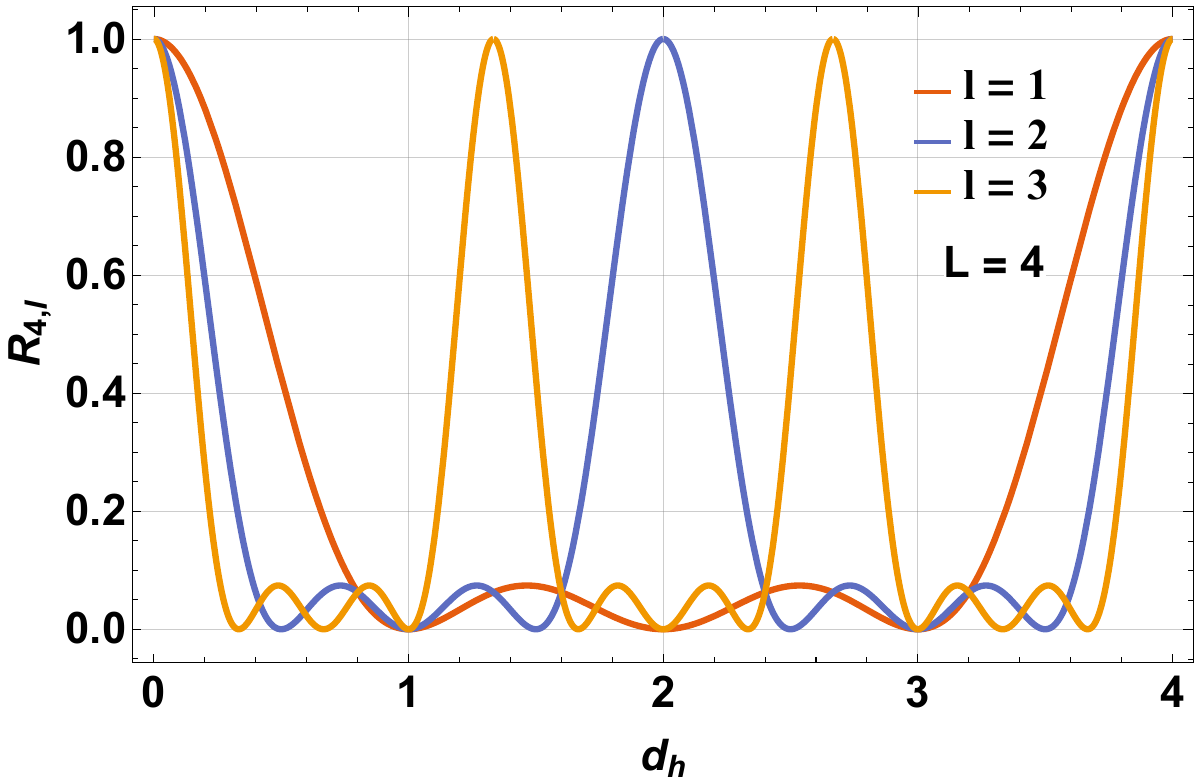}&
b)&\includegraphics[scale=0.40]{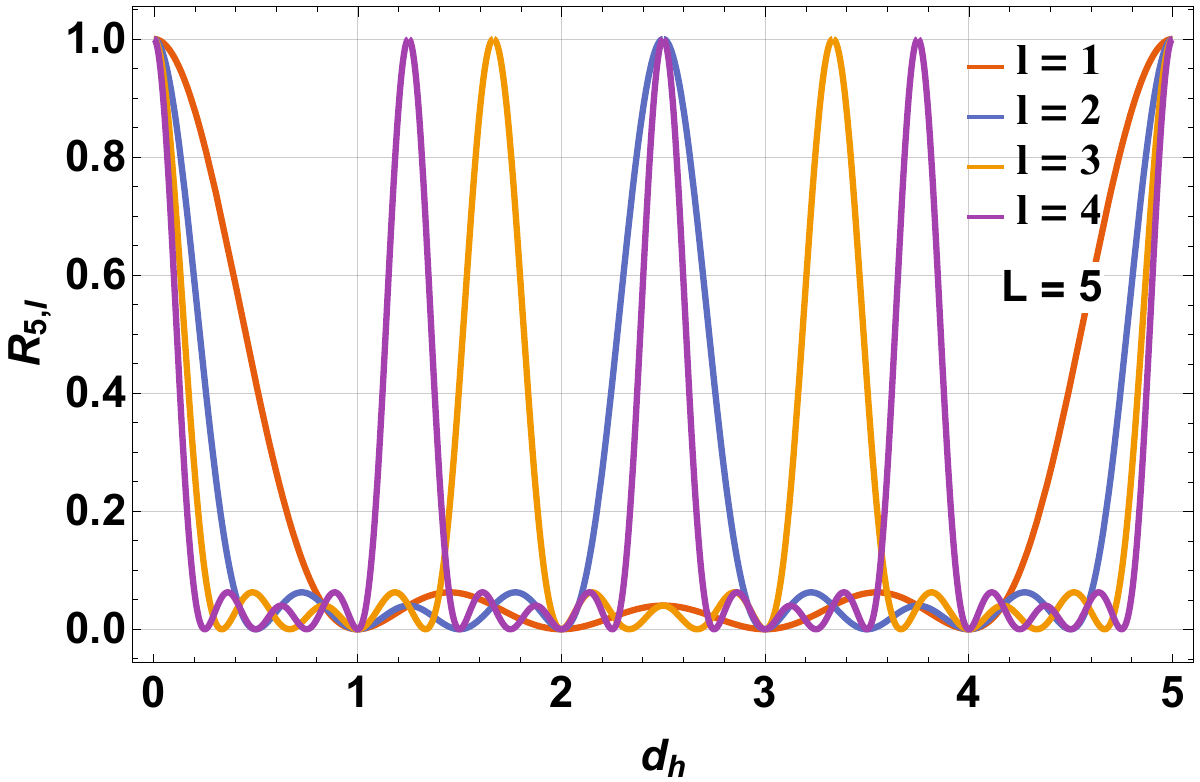}
\end{tabular}
\caption{\label{Fig6}\small The eigenvalue modulation function (a) $R_{4,\ell}(d_h)$, and (b)  $R_{5,\ell}(d_h)$ plotted v.s. $d_h$ for $\ell>0$ (eq.~\ref{lambdaohnebar}). Only the single period of each function is plotted.} 
\end{figure*}

As expected, at $\ell=0$, the leading eigenvalue is  $\lambda_0=1$, independently of the model's parameters. For  all other eigenvalues  we have $|\lambda_{\ell}| \leq |\bar{\lambda}_\ell| \leq 1$ due to the presence of the modulating function $R_{L,\ell}(d_h)$, bounded  from above and below, $0\le R_{L,\ell}(d_h)\le 1$. 
 At $d_h=0$ the modulating function equals $1$ identically, which returns us to the one-dimensional cat map chain. For $d_h=-1$ corresponding to the fully dual-unitary case, the function $R_{L,\ell}(d_h)$ equals zero identically  for $\ell>0$. From this observation, it follows that the correlation function $C(t)$ vanishes for the traceless observables, as it should be for a fully dual-unitary model.  Moreover, the function $R_{L,\ell}(d_h)$  and the correlation function $C(t)$ equal  zero  for almost all other integer values of $d_h$. Exceptions occur in cases where $L$ is a product of several prime numbers, for example, $L=p_1 p_2$ and $p_1,p_2\ne 1$. Here, the function $R_{L,\ell}(d_h)$ can attain its maximum
 value of $1$ at certain values of $d_h$ other than $0$.
For instance $R_{L=p_1p_2,\ell=p_1}(p_2)=R_{L=p_1p_2,\ell=p_2}(p_1)=1$. To demonstrate this features we plotted the functions $R_{L=4,\ell}(d_h)$, and $R_{L=4,\ell}(d_h)$ in fig.~\ref{Fig6}.

\subsection{The Kicked Ising Spin lattice}\label{SecVB}
In this part, we illustrate our results on the particular example of the minimal dimension model ($L=2$), the Kicked Ising  spin-1/2 lattice. This 
 model is  known to have the dual-unitary regime and has  served as a paradigm  in the field of  many-body quantum chaos, see ~\cite{prosenJt-2,AWGG16, BeKoPr18, BeKoPr19-1, GBAWG20}.  Although the model has primarily been investigated in one dimension ($D=1$), its extension to many-dimensional ($D>1$) lattices is straightforward~\cite{prosen2014}. The system evolution is governed by the Hamiltonians

\begin{eqnarray}
\label{spininter}
H_I &=& \sum_{n=1}^N\sum_{m=1}^M d_v\hat{\sigma}^{z}_{n,m}\hat{\sigma}^{z}_{n+1,m}+d_h\hat{\sigma}^{z}_{n,m}\hat{\sigma}^{z}_{n,m+1}+h\hat{\sigma}^{z}_{n,m}\nonumber\\
H_K &=& 
\sum_{n=1}^N\sum_{m=1}^M J\hat{\sigma}^{x}_{n,m}
\end{eqnarray}

where the operators $\hat{\sigma}^{\alpha}_{n,m}$  are the Pauli matrices with $\alpha=x,y,z$ (see appendix~\ref{AppC}), acting in the two-dimensional Hilbert space ($L=2$) of a single spinor with the lattice index $(m,n)$.  As everywhere above, we assume that $n=\overline{1,N}$, $m=\overline{1,M}$ and assume the cyclic boundary conditions, i.e. $\hat{\sigma}^{\alpha}_{N+1,m}\equiv \hat{\sigma}^{\alpha}_{1,m}$, $\hat{\sigma}^{\alpha}_{n,M+1}\equiv \hat{\sigma}^{\alpha}_{n,1}$.
To make the model partially dual-unitary we set $J=\pi/2, d_v=\pi/4$. With this choice
the matrix $u[g]$ has the form
\begin{equation}
    u[g]=\frac{1}{\sqrt{2}}\left(
    \begin{array}{cc}
      1   &  -\i \\
       -\i  &   1
    \end{array}
    \right),
\end{equation}
while the matrices $u[f_{v/h}]$ are given by
\begin{equation}
    u[f_{v/h}]=\frac{1}{\sqrt{2}}\left(
    \begin{array}{cc}
      e^{-\i(d_{v/h}+h_{v/h})}   &  e^{\i d_{v/h}} \\
       e^{\i d_{v/h}}  &  e^{-\i(d_{v/h}-h_{v/h})}
    \end{array}
    \right),
\end{equation}
where $h_h=h, h_v=0$
The function $g(s,s')=\frac{\pi}{4}\big((2s-3)(2s'-3)-1\big)$ and $f_{v/h}(s,s')=d_{v/h}(2s-3)(2s'-3)+h_{v/h}(s+s'-3)$. Note that  the matrices $u[f_v]$ and $u[g]$ coincide up to a constant phase factor $e^{\i\pi/4}$, which is irrelevant as far as the correlation function is concerned. In accordance with  the eq.~(\ref{TD2}), the resulting $4\times 4$ transfer matrix $\braket{\chi,\eta|\bm T|\chi',\eta'}$ in the basis  $\{\ket{1,1},\ket{2,2},\ket{1,2},\ket{2,1}\}$ has the  $2\times 2$ block structure
\begin{equation}\label{transIsing}
   T= \left(  
      \begin{array}{cc}
         \alpha E  & \beta E \\
         \beta E  & \alpha E
      \end{array} \right),    
\end{equation}
 \[
\alpha=\frac{1}{4} \big( 1+ \cos^2 2d_h\cos 2h \big),\quad \beta=\frac{1}{4}\big( 1- \cos^2 2d_h\cos 2h \big),  \]     
where $E$ is  the $2\times 2$ matrix  with the unit entries, i.e., $E_{i,j}=1$ for all $i,j$. The transfer matrix has two zero eigenvalues, the eigenvalue $\lambda_0=1$ with the  eigenvector  $\frac{1}{2}(1,1,1,1)$ and the eigenvalue
\begin{equation}\label{eig1Ising}
\lambda_1=\cos^2 2d_h\cos 2h,
\end{equation}
with the corresponding eigenvector  $\frac{1}{2}(-1,-1,1,1)$. 

By using the spectrum of $T$,  the correlation function $C(T)$ for the traceless operators $\q_i$ takes the form
\begin{widetext}
\begin{multline}\label{CTSpinModel}
C(T)=-\frac{1}{2} \lambda_1^{T-2}\cos^4 2d_h \bra{1}\q_1\ket{1} \bra{1}u^\dag[g]\q_6 u[g]\ket{1} \Big[\cos 2h (\bra{1}\q_2\ket{2} - \bra{2}\q_2\ket{1}) +\i \sin2h (\bra{1}\q_2\ket{2} + \bra{2}\q_2\ket{1})\Big]  \\ \times
\Big[\cos 2h (\bra{1}u^\dag[g]\q_5 u[g]\ket{2} - \bra{2}u^\dag[g]\q_5 u[g]\ket{1}) +\i \sin2h(\bra{1}u^\dag[g]\q_5 u[g]\ket{2} + \bra{2}u^\dag[g]\q_5 u[g]\ket{1})\Big] .
\end{multline}
\end{widetext}
It is instructive to calculate the correlation function for the operators $\q_\ell$ taken from the set of Pauli  matrices. Obviously, only the choice $\q_1=\hat{\sigma}^z$ and $\q_6=\hat{\sigma}^y$, $\q_2=\hat{\sigma}^x,\hat{\sigma}^y$, $\q_5=\hat{\sigma}^x, \hat{\sigma}^z$ corresponds to the non-zero correlation function, in total there are 4 combinations which lead to a non-trivial correlation function. The results of calculations are gathered in the appendix~\ref{AppC}.
 
Finally, consider the  transfer matrix (eq.~\ref{TTD2}) for correlation function of four-point supported operators in the kicked Ising spin lattice model. The matrix $\braket{\chi,\eta,\chi_1,\eta_1|\bm T|\chi',\eta',\chi'_1,\eta'_1}$  can be  written in  a block-hierarchical structured form
\begin{equation}
    T=\left( \begin{array}{cc}
      \begin{array}{cccc}
         \tilde{\alpha} E & \tilde{\beta}E &\tilde{\gamma}E&\tilde{\gamma}E\\
         \tilde{\beta}E  & \tilde{\alpha} E&\tilde{\gamma}E&\tilde{\gamma}E
         \\
\tilde{\gamma}E&\tilde{\gamma}E &\tilde{\alpha} E  & \tilde{\beta}E \\
 \tilde{\gamma}E&\tilde{\gamma}E        &\tilde{\beta} E & \tilde{\alpha}E
      \end{array}
    \end{array}\right),
\end{equation}
where
\begin{eqnarray*}
\tilde{\alpha}&=&\frac{1}{8}\big(4\cos^4 h\cos^2 2d_h+\sin^22d_h\big),\\ \tilde{\beta}&=&\frac{1}{8}\big(4\sin^4 h\cos^2 2d_h+\sin^22d_h\big)\\
\tilde{\gamma}&=&\frac{1}{32}\big(3-\cos4 h-2\cos^2 2h\cos4d_h\big).
\end{eqnarray*}
 This transfer matrix possesses four non-zero eigenvalues: the eigenvalue $\lambda_0=1$, two degenerated eigenvalues $\lambda_{1,2}=\cos^22d_h\cos2h$ (coincide with the eigenvalue $\lambda_1$ for the two-point transfer matrix, eq.~\ref{eig1Ising}), and the eigenvalue $\lambda_3=\cos^22d_h\cos^22h$. Note that $\lambda_{1,2}$ are identical to the second-largest eigenvalue (eq.~\ref{eig1Ising}) of the  transfer matrix (\ref{transIsing}). This implies that the decay rate of the correlations between operators with the two-point  and four-point supports coincide.

\section{Conclusion}\label{Sec8}
In the current research, we have explored two-dimensional lattice models featuring partial spatiotemporal symmetry. The study revealed that  for partially dual-unitary models, non-trivial  correlations exist   along  the  light cone edges in the  space-time grid. We have expressed these correlations through the expectation values of  powers of a low-dimensional transfer matrix. On the other hand, fully dual-unitary models exhibit ultra-local correlations that completely vanish after a finite time. These findings corroborate earlier observations \cite{BeKoPr19-1} indicating that (fully) dual-unitary models constitute a maximally chaotic class of systems.

 As an illustration, we applied these findings to the coupled quantum cat maps and the kicked Ising spin-lattice. For these models we have derived an explicit formula for the spectrum of the transfer operator $T$, enabling us to determine decay rates of correlation functions for operators with two-point and four-point supports.
Remarkably, the second-largest eigenvalues of $T$ attain a simple structure - it is provided here by the second-largest eigenvalue  of the transfer operator for the corresponding one-dimensional model, multiplied by a factor. The absolute value of this factor depends on the coupling in non-dual directions and is bounded from above by one.
This demonstrates that the inclusion of an extra spatial dimension generally enhances the decay rate of the correlation function.

It is worth noticing that the above results can be straightforwardly extended to lattice models with arbitrary dimensions $D>2$. Assuming that dual-unitarity holds for at least two spatial dimensions (e.g., 1, 2), we can conclude that the correlation function can take non-trivial values
\begin{equation}
C(\bm r, t)\neq0, \qquad  \bm r =(n_1,n_2, \dots n_D)
\end{equation}
if and only if  the following three inequalities hold:
\begin{eqnarray}\label{ineq1}
|t|&\geq& \sum_{i=1}^D|n_i|, \\ \label{ineq2} |n_1|&\geq& |t|+\sum_{i\neq 1}^D|n_i|\\
|n_2|&\geq& |t|+\sum_{i\neq 2}^D |n_i|. 
\end{eqnarray}
As these are satisfied only at a single point, $t=0,\bm r=0$,   all correlations in this case  are ultra-local, meaning they vanish identically after a finite time.  If, however, the  dual-unitarity holds solely for a single spatial dimension, then only the first two inequalities are satisfied. In this case, the non-trivial correlations emerge along the line $|t|=|n_1|$, $n_i=0, i\neq 1$. As in the two-dimensional case,  $C(\bm r, t)$ can be expressed through a transfer operator, whose dimension is determined by the size of the local operator's support.   
    
\acknowledgments
This work was supported by the Israel Science Foundation (ISF) grant
No. 2089/19. Two of us (VO and BG) thank the University of
Duisburg-Essen for hospitality during their visit in February of 2023.


\bibliographystyle{ieeetr}

\begin{thebibliography}{50}

\bibitem{haake}
F.~Haake, {\em {Quantum Signatures of Chaos}}.
\newblock {Springer Series in Synergetics}, Springer, 3~ed., 2010.

\bibitem{guhr}
T.~Guhr, A.~M{\"u}ller-Groeling, and H.~A. Weidenm{\"u}ller, ``{Random-matrix
  theories in quantum physics: common concepts},'' {\em Physics Reports},
  vol.~299, no.~4--6, pp.~189--425, 1998.

\bibitem{fisher2023random}
M.~P. Fisher, V.~Khemani, A.~Nahum, and S.~Vijay, ``Random quantum circuits,''
  {\em Annual Review of Condensed Matter Physics}, vol.~14, pp.~335--379, 2023.

\bibitem{GutOsi15}
B.~Gutkin and V.~Osipov, ``Classical foundations of many-particle quantum
  chaos,'' {\em Nonlinearity}, vol.~29, pp.~325--356, 2016.

\bibitem{GHJSC16}
B.~Gutkin, P.~Cvitanovi{\'{c}}, R.~Jafari, A.~K. Saremi, and L.~Han, ``Linear
  encoding of the spatiotemporal cat,'' {\em Nonlinearity}, vol.~34,
  pp.~2800--2836, may 2021.

\bibitem{LiangCvit20022}
H.~Liang and P.~Cvitanović, ``A chaotic lattice field theory in one
  dimension,'' {\em Journal of Physics A: Mathematical and Theoretical},
  vol.~55, p.~304002, jul 2022.

\bibitem{Fouxon_Gutkin_2022}
I.~Fouxon and B.~Gutkin, ``Local correlations in coupled cat maps with
  space-time duality,'' {\em Journal of Physics A: Mathematical and
  Theoretical}, vol.~55, p.~504004, dec 2022.

\bibitem{AWGG16}
M.~Akila, D.~Waltner, B.~Gutkin, and T.~Guhr, ``Particle-time duality in the
  kicked {Ising} spin chain,'' {\em J. Phys. A}, vol.~49, p.~375101, 2016.

\bibitem{BeKoPr18}
B.~Bertini, P.~Kos, and T.~Prosen, ``Exact spectral form factor in a minimal
  model of many-body quantum chaos,'' {\em Phys. Rev. Lett.}, vol.~121,
  p.~264101, 2018.

\bibitem{BeKoPr19-1}
B.~Bertini, P.~Kos, and T.~Prosen, ``Entanglement spreading in a minimal model
  of maximal many-body quantum chaos,'' {\em Phys. Rev. X}, vol.~9, p.~021033,
  2019.

\bibitem{GBAWG20}
B.~Gutkin, P.~Braun, M.~Akila, D.~Waltner, and T.~Guhr, ``Exact local
  correlations in kicked chains,'' {\em Phys. Rev. B}, vol.~102, p.~174307, Nov
  2020.

\bibitem{BeKoPr19-4}
B.~Bertini, P.~Kos, and T.~Prosen, ``Exact correlation functions for
  dual-unitary lattice models in $1+1$ dimensions,'' {\em Phys. Rev. Lett.},
  vol.~123, p.~210601, Nov 2019.

\bibitem{Arul19}
S.~A. Rather, S.~Aravinda, and A.~Lakshminarayan, ``Creating ensembles of dual
  unitary and maximally entangling quantum evolutions,'' {\em Phys. Rev.
  Lett.}, vol.~125, p.~070501, Aug 2020.

\bibitem{GopLam19}
S.~Gopalakrishnan and A.~Lamacraft, ``Unitary circuits of finite depth and
  infinite width from quantum channels,'' {\em Phys. Rev. B}, vol.~100,
  p.~064309, 2019.

\bibitem{LakshPal2018}
R.~Pal and A.~Lakshminarayan, ``Entangling power of time-evolution operators in
  integrable and nonintegrable many-body systems,'' {\em Phys. Rev. B},
  vol.~98, p.~174304, Nov 2018.

\bibitem{BWAGG19}
P.~Braun, D.~Waltner, M.~Akila, B.~Gutkin, and T.~Guhr, ``Transition from
  quantum chaos to localization in spin chains,'' {\em Phys. Rev. E}, vol.~101,
  p.~052201, May 2020.

\bibitem{BeKoPrPi19}
L.~Piroli, B.~Bertini, J.~I. Cirac, and T.~c.~v. Prosen, ``Exact dynamics in
  dual-unitary quantum circuits,'' {\em Phys. Rev. B}, vol.~101, p.~094304, Mar
  2020.

\bibitem{BeKoPr2019operator}
B.~Bertini, P.~Kos, and T.~Prosen, ``{Operator Entanglement in Local Quantum
  Circuits I: Chaotic Dual-Unitary Circuits},'' {\em SciPost Phys.}, vol.~8,
  p.~067, 2020.

\bibitem{KPBBPT_2021}
P.~Kos, B.~Bertini, and T.~c.~v. Prosen, ``Correlations in perturbed
  dual-unitary circuits: Efficient path-integral formula,'' {\em Phys. Rev. X},
  vol.~11, p.~011022, Feb 2021.

\bibitem{zhou2019entanglement}
T.~Zhou and A.~Nahum, ``Entanglement membrane in chaotic many-body systems,''
  {\em Phys. Rev. X}, vol.~10, p.~031066, Sep 2020.

\bibitem{AVAN2016}
J.~Avan, V.~Caudrelier, A.~Doikou, and A.~Kundu, ``Lagrangian and hamiltonian
  structures in an integrable hierarchy and space-time duality,'' {\em Nuclear
  Physics B}, vol.~902, pp.~415 -- 439, 2016.

\bibitem{Karl15}
D.~Goyeneche, D.~Alsina, J.~I. Latorre, A.~Riera, and K.~\ifmmode~\dot{Z}\else
  \.{Z}\fi{}yczkowski, ``Absolutely maximally entangled states, combinatorial
  designs, and multiunitary matrices,'' {\em Phys. Rev. A}, vol.~92, p.~032316,
  Sep 2015.

\bibitem{Arul2021}
S.~Aravinda, S.~A. Rather, and A.~Lakshminarayan, ``From dual-unitary to
  quantum bernoulli circuits: Role of the entangling power in constructing a
  quantum ergodic hierarchy,'' {\em Phys. Rev. Research}, vol.~3, p.~043034,
  Oct 2021.

\bibitem{PhysRevResearch.3.023118}
A.~Chan, A.~De~Luca, and J.~T. Chalker, ``Spectral lyapunov exponents in
  chaotic and localized many-body quantum systems,'' {\em Phys. Rev. Res.},
  vol.~3, p.~023118, May 2021.

\bibitem{CL20}
P.~W. Claeys and A.~Lamacraft, ``Maximum velocity quantum circuits,'' {\em
  Phys. Rev. Research}, vol.~2, p.~033032, Jul 2020.

\bibitem{borsi2022construction}
M.~Borsi and B.~Pozsgay, ``Construction and the ergodicity properties of dual
  unitary quantum circuits,'' {\em Physical Review B}, vol.~106, no.~1,
  p.~014302, 2022.

\bibitem{ippoliti2021postselection}
M.~Ippoliti and V.~Khemani, ``Postselection-free entanglement dynamics via
  spacetime duality,'' {\em Physical Review Letters}, vol.~126, no.~6,
  p.~060501, 2021.

\bibitem{claeys2022exact}
P.~W. Claeys, M.~Henry, J.~Vicary, and A.~Lamacraft, ``Exact dynamics in
  dual-unitary quantum circuits with projective measurements,'' {\em Physical
  Review Research}, vol.~4, no.~4, p.~043212, 2022.

\bibitem{krajnik2020kardar}
{\v{Z}}.~Krajnik and T.~Prosen, ``Kardar--parisi--zhang physics in integrable
  rotationally symmetric dynamics on discrete space--time lattice,'' {\em
  Journal of Statistical Physics}, vol.~179, no.~1, pp.~110--130, 2020.

\bibitem{lu2021spacetime}
T.-C. Lu and T.~Grover, ``Spacetime duality between localization transitions
  and measurement-induced transitions,'' {\em PRX Quantum}, vol.~2, no.~4,
  p.~040319, 2021.

\bibitem{prosen2021many}
T.~Prosen, ``Many-body quantum chaos and dual-unitarity round-a-face,'' {\em
  Chaos: An Interdisciplinary Journal of Nonlinear Science}, vol.~31, no.~9,
  2021.

\bibitem{piroli2020exact}
L.~Piroli, B.~Bertini, J.~I. Cirac, and T.~Prosen, ``Exact dynamics in
  dual-unitary quantum circuits,'' {\em Physical Review B}, vol.~101, no.~9,
  p.~094304, 2020.

\bibitem{lerose2021influence}
A.~Lerose, M.~Sonner, and D.~A. Abanin, ``Influence matrix approach to
  many-body floquet dynamics,'' {\em Physical Review X}, vol.~11, no.~2,
  p.~021040, 2021.

\bibitem{ippoliti2022fractal}
M.~Ippoliti, T.~Rakovszky, and V.~Khemani, ``Fractal, logarithmic, and
  volume-law entangled nonthermal steady states via spacetime duality,'' {\em
  Physical Review X}, vol.~12, no.~1, p.~011045, 2022.

\bibitem{fritzsch2021eigenstate}
F.~Fritzsch and T.~Prosen, ``Eigenstate thermalization in dual-unitary quantum
  circuits: Asymptotics of spectral functions,'' {\em Physical Review E},
  vol.~103, no.~6, p.~062133, 2021.

\bibitem{suzuki2022computational}
R.~Suzuki, K.~Mitarai, and K.~Fujii, ``Computational power of one-and
  two-dimensional dual-unitary quantum circuits,'' {\em Quantum}, vol.~6,
  p.~631, 2022.

\bibitem{flack2020statistics}
A.~Flack, B.~Bertini, and T.~Prosen, ``Statistics of the spectral form factor
  in the self-dual kicked ising model,'' {\em Physical Review Research},
  vol.~2, no.~4, p.~043403, 2020.

\bibitem{bertini2021random}
B.~Bertini, P.~Kos, and T.~Prosen, ``Random matrix spectral form factor of
  dual-unitary quantum circuits,'' {\em Communications in Mathematical
  Physics}, vol.~387, no.~1, pp.~597--620, 2021.

\bibitem{jonay2021triunitary}
C.~Jonay, V.~Khemani, and M.~Ippoliti, ``Triunitary quantum circuits,'' {\em
  Physical Review Research}, vol.~3, no.~4, p.~043046, 2021.

\bibitem{AWGBG16}
M.~Akila, D.~Waltner, B.~Gutkin, P.~Braun, and T.~Guhr, ``Semiclassical
  identification of periodic orbits in a quantum many-body system,'' {\em Phys.
  Rev. Lett.}, vol.~118, p.~164101, 2017.

\bibitem{AGBWG18}
M.~Akila, B.~Gutkin, P.~Braun, D.~Waltner, and T.~Guhr, ``Semiclassical
  prediction of large spectral fluctuations in interacting kicked spin
  chains,'' {\em Ann. Phys.}, vol.~389, pp.~250--282, 2018.

\bibitem{ArnoldBook}
V.~Arnold and A.~Avez, {\em Ergodic problems of classical mechanics}.
\newblock Redwood City (Calif.), Addison-Wesley., 1989.

\bibitem{HannayBerry1980}
J.~Hannay and M.~Berry, ``Quantization of linear maps on a torus-fresnel
  diffraction by a periodic grating,'' {\em Physica D: Nonlinear Phenomena},
  vol.~1, no.~3, pp.~267--290, 1980.

\bibitem{Rivas_2000}
A.~M.~F. Rivas, M.~Saraceno, and A.~M.~O. de~Almeida, ``Quantization of
  multidimensional cat maps,'' {\em Nonlinearity}, vol.~13, pp.~341--376, jan
  2000.

\bibitem{prosenJt-2}
T.~Prosen, ``{Exact Time-Correlation Functions of Quantum Ising Chain in a
  Kicking Transversal Magnetic Field: Spectral Analysis of the Adjoint
  Propagator in Heisenberg Picture},'' {\em Progress of Theoretical Physics
  Supplement}, vol.~139, p.~191, 2000.

\bibitem{prosen2014}
C.~Pineda, T.~Prosen, and E.~Villase{\~n}or, ``{Two dimensional kicked quantum
  Ising model: dynamical phase transitions},'' {\em New Journal of Physics},
  vol.~16, no.~12, p.~123044, 2014.

\end{thebibliography}

\appendix

\section{The spin structures obtained by application of the contraction rules}\label{AppB}
\begin{figure*}
\begin{tabular}{lccccc}
a)&\includegraphics[scale=0.25]{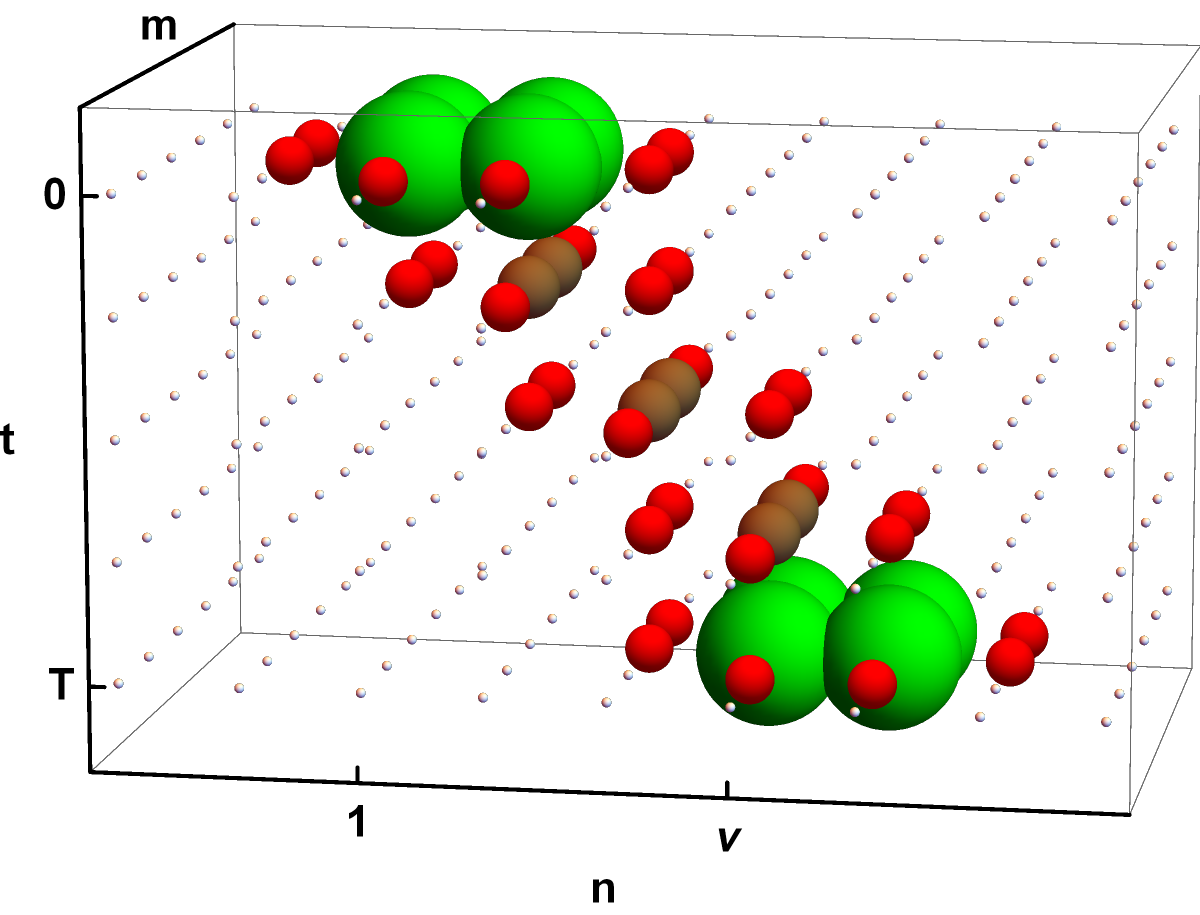}&
b)&\includegraphics[scale=0.25]{Ex2.pdf}&
c)&\includegraphics[scale=0.25]{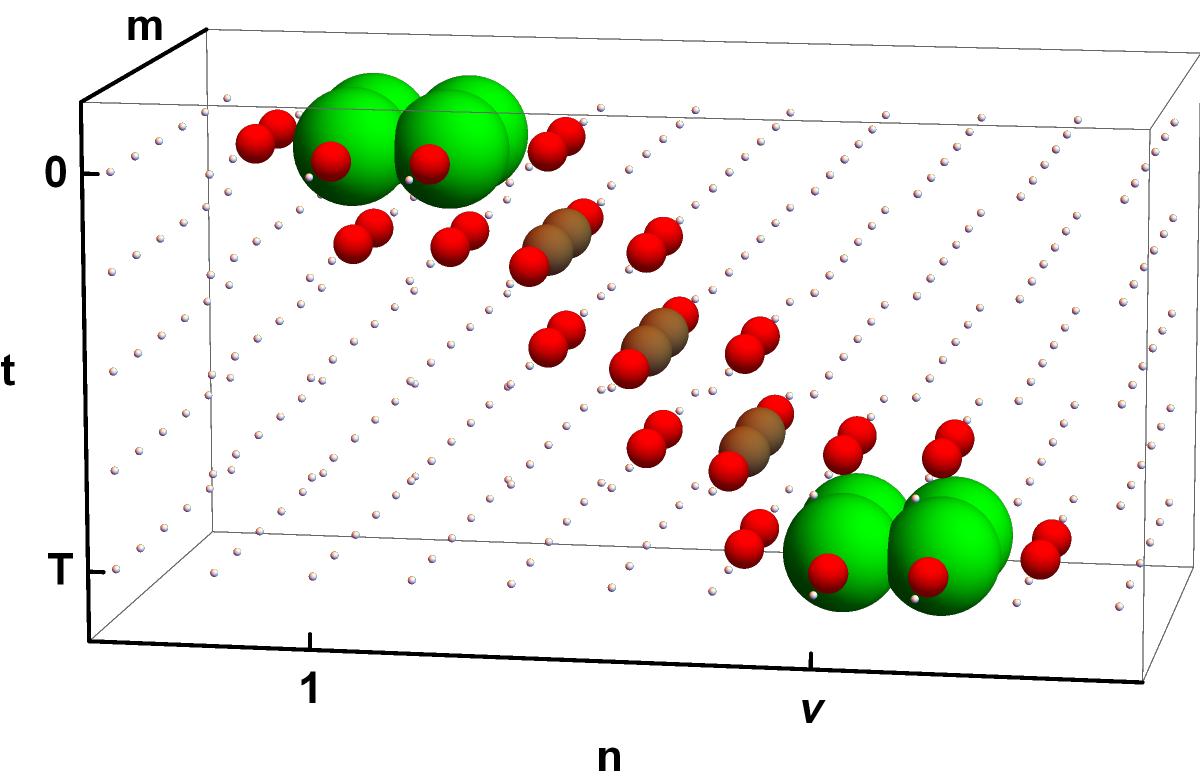}
\end{tabular}
\caption{\label{Fig1AppB}\small The spin structures generated after the application of the contraction rules in the temporal and the horizontal directions at $T=4$ ($M,N\gg T$) and  (a) $\nu=4$, (b) $\nu=5$, (c) $\nu=6$.} 
\end{figure*}

Multiple applications of the contraction rules formulated in section~\ref{Sec3} to the original spin structure in the case of the two-dimensional partially dual map (the horizontal direction is dual to the time direction) and for $T\gg M, N$ generates a number of non-trivial structures. In the generic case of the eight-point correlation function, there are three (up to the mirror transformation $n\to N-n$) non-trivial final structures (fig.~\ref{Fig1AppB}) taking place at $\nu=T, T+1,T+2$ (see eqs.~\ref{barSigma}). Further analysis shows that only the case $\nu=T+1$ corresponds to the non-trivial correlation function. Moreover, the structure similar to the one shown in fig.~\ref{Fig1AppB}b reduces to those in fig.~\ref{Fig5}b. These conclusions follow from the explicit summation of the correlated spins (red balls) in the vicinity of the boundaries (green balls).

On the upper boundary of the structure shown in fig.~\ref{Fig1AppB}a, the summation over the correlated spins $s_{N,1,0}$, $s_{N,2,0}$, $s_{3,1,0}$, $s_{3,2,0}$ has to be done according to the following scheme (only the spins with $m=1$ are shown)
\begin{widetext}
\begin{equation}\label{schemeAppB}
\begin{tikzcd}  
t=0&s_{N,1,0}  \arrow[r,"f_v"] &\bra{\bar{s}_{1,1,0}}\q_1\ket{\ubar{s}_{1,1,0}} \arrow[r,"f_v"]& \bra{\bar{s}_{2,1,0}}\q_2\ket{\ubar{s}_{2,1,0}}\arrow[r,"f_v"] & s_{3,1,0}\\
t=1&&s_{1,1,1}\arrow[u,"g"] \arrow[r,"f_v"] &  (\bar{s}_{2,1,1},\ubar{s}_{2,1,1}) \arrow[r,"f_v"] \arrow[u,"g"] &  \dots
\end{tikzcd}
\end{equation}  
\end{widetext}
The arrows on this scheme show the order of indexes in the corresponding functions ($f_v$ of $g$), from left to right. The horizontal interactions are not shown, while the presence of the horizontal interactions after summation results in the correlation of the spins $\bar{s}_{1,1,0}$ and $\ubar{s}_{1,1,0}$. Thus, on the next turn, summation over the spin $s_{1,1,1}$ leads to the correlation of $\bar{s}_{2,1,1}$ and $\ubar{s}_{2,1,1}$, which cuts the spin-bridge and the correlation function becomes trivial (the same arguing works for the spins with $m=2$).
  
The spin structure in fig.~\ref{Fig1AppB}c has to be analysed starting from the bottom. To formulate the contraction rules in a symmetric manner we have introduced the additional matrices $U_I$ into $\Phi(\bar{\bm s}_T,\ubar{\bm s}_T)$, so that the product $\Phi(\bar{\bm s}_T,\ubar{\bm s}_T)\bar{T}_I(\bar{\bm s}_T,\ubar{\bm s}_T)$ reduces to the scalar product $\bra{\bar{\bm s}_T}\ubar{\Sigma}\ket{\ubar{\bm s}_T}$, where all spins on the level $t=T$ are correlated and does not interact horizontally. Therefore summation over the spins $s_{T+1,1,T}$ and $s_{T+1,2,T}$ results in the correlation of the spin pairs $\bar{s}_{T+1,1,T-1}$,  $\ubar{s}_{T+1,1,T-1}$ and  $\bar{s}_{T+1,2,T-1}$,  $\ubar{s}_{T+1,2,T-1}$, which again breaks the spin-bridge. The very same arguments allow us to reduce the spin structure in fig.~\ref{Fig1AppB}b to obtain the one in  fig.~\ref{Fig5}b.

\begin{figure*}
\begin{tabular}{lccccc}
a)&\includegraphics[scale=0.25]{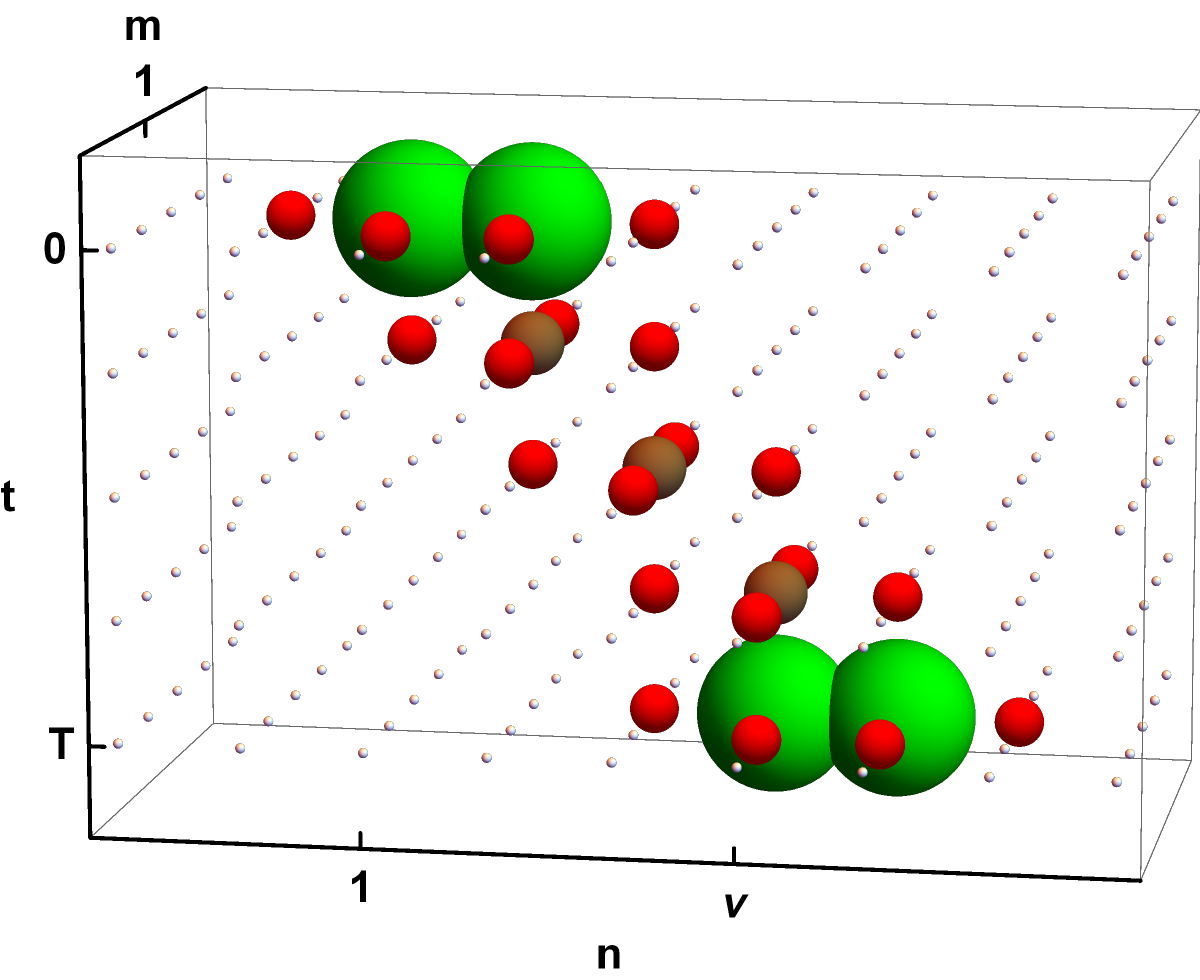}&
b)&\includegraphics[scale=0.25]{Ex8.pdf}&
c)&\includegraphics[scale=0.25]{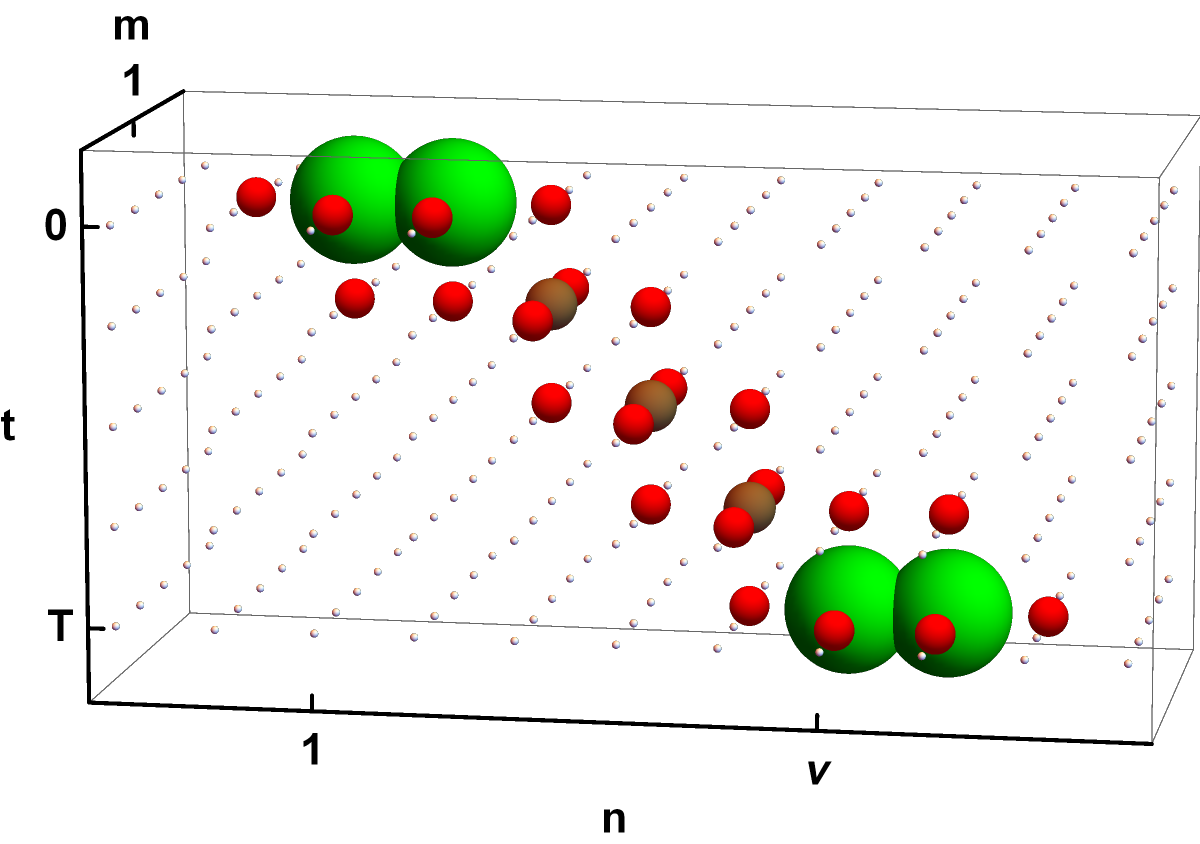}\\
e)&\includegraphics[scale=0.25]{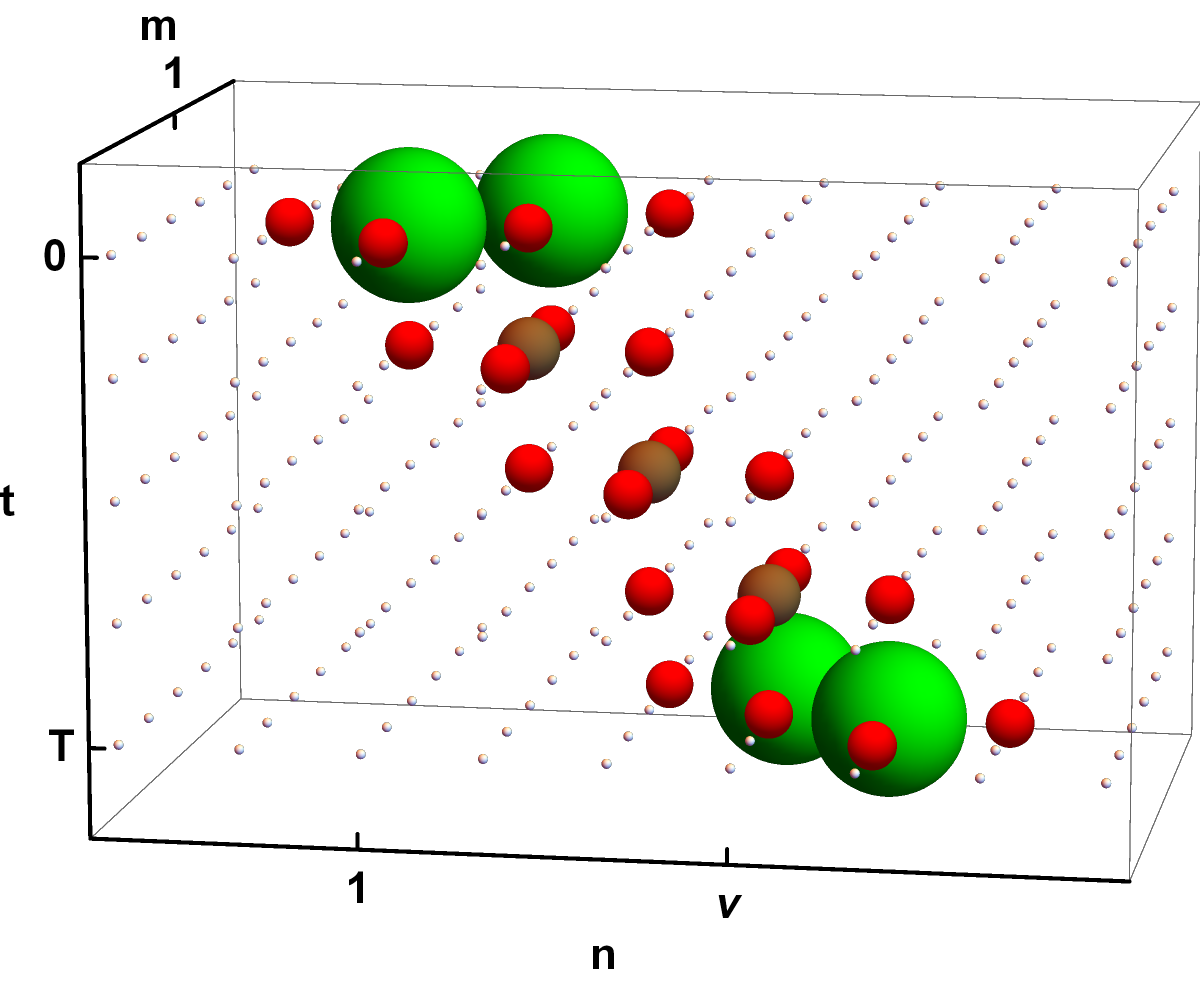}&
d)&\includegraphics[scale=0.25]{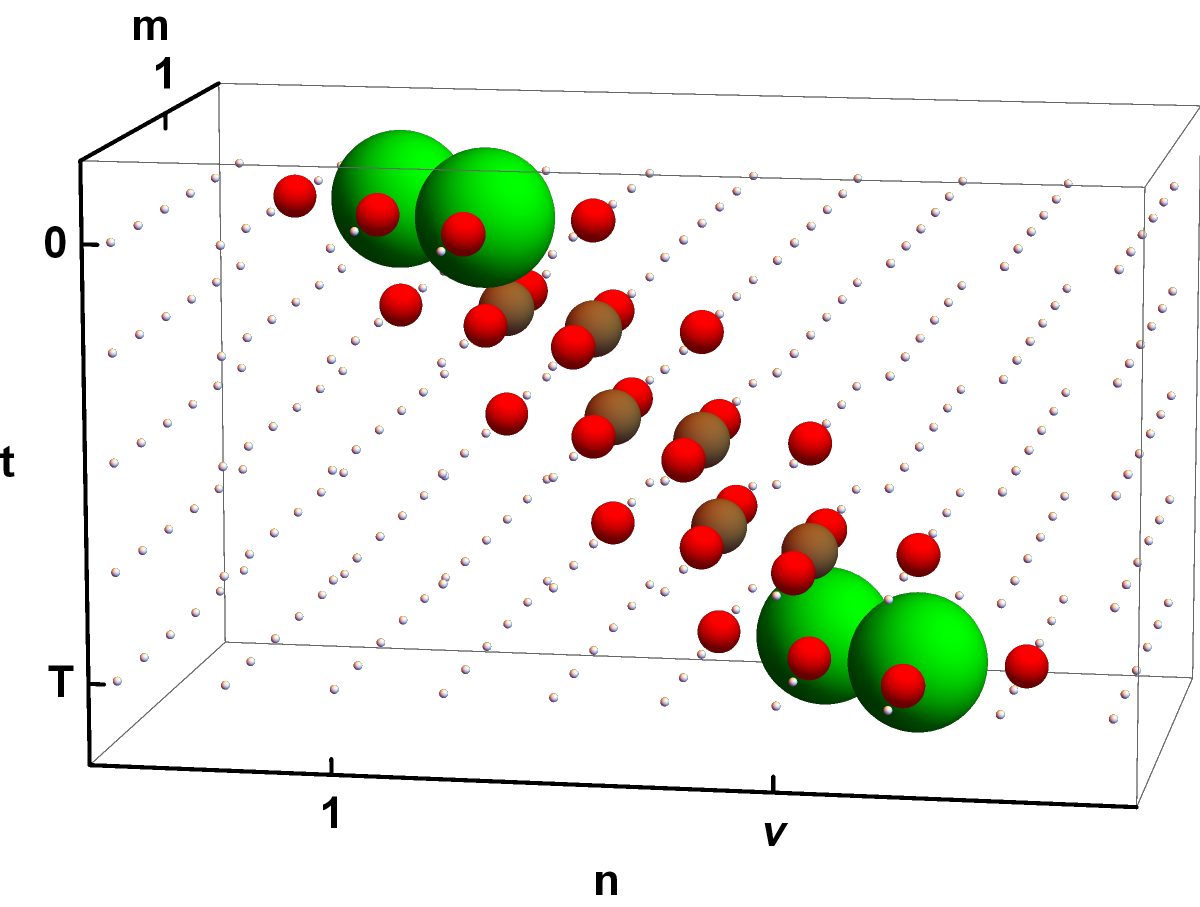}&
f)&\includegraphics[scale=0.25]{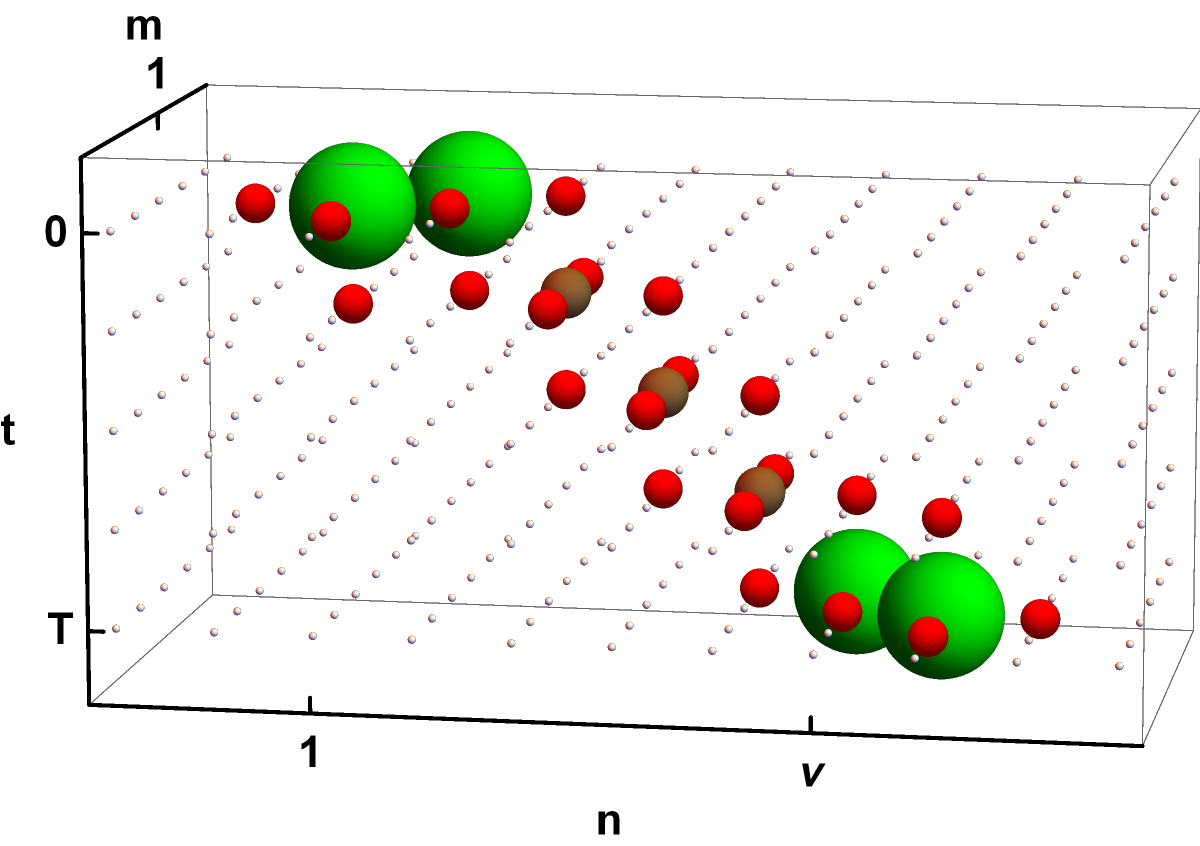}
\end{tabular}
\caption{\label{Fig3AppB}\small The spin structures generated after the application of the contraction rules in the temporal and the horizontal directions at $T=4$ ($M,N\gg T$) when only four from eight operators $q_\ell$ are different from $\1$. The mutual positions of the operators $q_\ell$ are different at each plot and (a, d) $\nu=4$, (b, d) $\nu=5$, (c, f) $\nu=6$.} 
\end{figure*}

For completeness, we also plotted the structures obtained after the application of the contraction rules for the case corresponding to the four-point correlation function with various mutual positions of the operators $\q_\ell$, see fig.~\ref{Fig3AppB}. Additional analysis shows that only one of them shown in  fig.~\ref{Fig3AppB}b generates a non-trivial correlation function. It can be reduced to the structure in fig.~\ref{Fig4}b.

\section{Explicit expressions for the correlation function in the spin-chain model (section~\ref{SecVB}) }\label{AppC}
The correlation function calculated in eq.~(\ref{CTSpinModel}) take non-zero values for particular choices of the operators $\q_i$: $\q_1=\hat{\sigma}^z$, $\q_6=\hat{\sigma}^y$, $\q_2=\hat{\sigma}^x,\hat{\sigma}^y$, $\q_5=\hat{\sigma}^x, \hat{\sigma}^z$, totally 4 combinations. We use the standard definition of the Pauli matrices,
\begin{equation}
    \hat{\sigma}^z=\left(\begin{array}{cc}
        1 & 0 \\
        0 & -1
    \end{array}\right),\quad  \hat{\sigma}^x=\left(\begin{array}{cc}
        0 & 1 \\
        1 & 0
    \end{array}\right),\quad  \hat{\sigma}^y=\left(\begin{array}{cc}
        0 & -\i \\
        \i & 0
    \end{array}\right).
\end{equation}
The correlation function is
\begin{equation}
C(T)=- 2\lambda_1^{T-2}\cos^4 2d_h \times \big(\mathrm{I}\big)\times\big(\mathrm{II}\big),
\end{equation}
where the factors $\big(\mathrm{I}\big)$, $\big(\mathrm{II}\big)$ take the following values:
\begin{equation}
\big(\mathrm{I}\big)=\begin{cases}   \sin2h& \q_2=\hat{\sigma}^x;\\
-  \cos 2h &\q_2=\hat{\sigma}^y;
\end{cases}
\end{equation}
\begin{equation}
\big(\mathrm{II}\big)=\begin{cases}    \sin2h & \q_5=\hat{\sigma}^x;\\
-  \cos 2h &\q_5=\hat{\sigma}^z.
\end{cases}
\end{equation}

\end{document}